\documentclass[aps,pre,twocolumn,showpacs]{revtex4}
\usepackage{graphicx}
\usepackage{amsmath,amscd,amssymb}
\usepackage{color}

\newcommand{\be}{\begin{equation}}
\newcommand{\ee}{\end{equation}}
\newcommand{\bea}{\begin{eqnarray}}
\newcommand{\eea}{\end{eqnarray}}
\newcommand{\lan}{\left\langle}
\newcommand{\ran}{\right\rangle}
\newcommand{\br}{\mathbf{r}}
\newcommand{\ba}{\mathbf{a}}

\newcommand{\bb}{\mathbf{b}}

\newcommand{\bu}{\mathbf{v}}

\newcommand{\bx}{\mathbf{x}}
\newcommand{\by}{\mathbf{y}}
\newcommand{\bo}{\mathbf{\Omega}_k}
\newcommand{\bom}{\mathbf{\Omega}}
\newcommand{\bk}{\mathbf{k}}
\newcommand{\bv}{\mathbf{u}}
\newcommand{\e}{\varepsilon}

\newcommand{\eo}{\varepsilon_0}

\newcommand{\ew}{\varepsilon_w}

\newcommand{\te}{\tilde{\epsilon}}

\newcommand{\tchi}{\tilde{\chi}}

\newcommand{\bz}{\bar{z}}

\newcommand{\tk}{\tilde{k}}

\begin{document}

\title{Microscopic formulation of non-local electrostatics in polar liquids embedding polarizable ions}

\author{Sahin Buyukdagli$^{1}$\footnote{email:~\texttt{sahin\_buyukdagli@yahoo.fr}} and T. Ala-Nissila$^{1,2}$\footnote{email:~\texttt{Tapio.Ala-Nissila@aalto.fi}}}
\affiliation{$^{1}$Department of Applied Physics and COMP center of Excellence, Aalto University School of Science, P.O. Box 11000, FI-00076 Aalto, Espoo, Finland\\
$^{2}$Department of Physics, Brown University, Providence, Box 1843, RI 02912-1843, U.S.A.}
\date{\today}

\begin{abstract}
Non-local electrostatic interactions associated with the finite solvent size and ion polarizability are investigated within the mean-field linear response theory.  To this end, we introduce a field theoretic model of a polar liquid composed of linear multipole solvent molecules and embedding polarizable ions modeled as Drude oscillators. \textcolor{black}{Unlike previous dipolar Poisson-Boltzmann formulations treating the solvent molecules as point dipoles, our model is able to qualitatively reproduce the non-local dielectric response behavior of polar liquids observed in Molecular Dynamics simulations and Atomic Force Microscope experiments for water solvent at charged interfaces.} The present theory explains the formation of the associated interfacial hydration layers in terms of a cooperative dipolar response mechanism driven by the reaction of the solvent molecules to their own polarization field. We also incorporate into the theory the relative multipole/dipole moments of water molecules obtained from quantum mechanical calculations, and show that the multipolar contributions to the dielectric permittivity are largely dominated by the dipolar one. We find that this stems from the mutual cancellation of the first two interfacial hydration layers of opposite net charge for multipolar liquids. Within the same non-local dielectric response theory, we show that the induced ion polarizability reverses the interfacial ion density trends predicted by the Poisson-Boltzmann theory, resulting in a surface affinity of coions and exclusion of counterions. The results indicate that the consideration of the discrete charge composition of solvent molecules and ions is the key step towards a microscopic understanding of non-local electrostatic effects in polar solvents.
\end{abstract}

\maketitle
\section{Introduction}

The precise determination of electrostatic interactions in the vicinity of charged molecules in water solvent is one of the biggest challenges in colloidal sciences. From the performance of energy storage devices~\cite{rev1,prlnetz,epl} and water purification membranes~\cite{yar,Lang} to the solubility of salt ions~\cite{rev2} and polyelectrolytes in water~\cite{rev3}, a wide variety of electrostatically driven nanoscale processes depend on the electrostatic potential behavior close to charged substrates. Water that mediates these electrostatic interactions being a strongly polar liquid, the evaluation of the electrostatic potential requires in turn a proper insight into the dielectric response of solvent molecules to the charged sources. Furthermore, the ordering of solvent molecules at charged mica surfaces revealed by Atomic Force Microscope (AFM) experiments~\cite{expdiel} and Molecular Dynamics (MD) simulations~\cite{Hans,prlnetz,langnetz} indicates that the electrostatic interactions in real systems are non-local. Therefore, a consistent formulation of non-local electrostatics is needed to understand nanoscale phenomena.

Our limited understanding of the dielectric response of water mainly stems from the lack of a microscopic theory able to map from the molecular details of the solvent to experimentally accessible macroscopic observables. In particular, dielectric continuum theories that bypass the charge composition of solvent molecules are unable to account for their coordinated behavior in the presence of charged sources. This approximation resulting in a uniform dielectric response characterized by a constant dielectric permittivity $\e(\br)=\e_w$  has severe drawbacks. For example, in bulk polar liquids, the local Born theory that cannot account for the dielectric void around ions is known to strongly overestimate ionic solvation energies~\cite{blossey1}.

Macroscopic theories of non-local electrostatics based on a phenomenological dielectric permittivity function $\e(\br,\br')$ and providing better agreement with experiments have been proposed over the last three decades. Among several efforts in this direction, one can mention the seminal works from A.A. Kornyshev et al. that dealt with non-local effects on the solvation of ions in bulk liquids~\cite{Kor1,Kor3} and the charge storage ability of metallic capacitors~\cite{Kor2}. Different formulations incorporating dipolar correlation effects in a coarse-grained way have been also developed in order to improve over the local Born theory in bulk~\cite{blossey1} and confined solvents~\cite{blossey2}.

In inhomogeneous liquids, the Poisson-Boltzmann (PB) formalism based on the dielectric continuum approximation fails as well to describe the interfacial dipolar ordering effect observed in experiments and simulations. The first dipolar Poisson-Boltzmann (DPB) approach able to account for the electrostatics of solvent molecules was introduced in Ref.~\cite{dunyuk}. The excluded volume of ions and solvent molecules was later incorporated to this formalism at the mean-field (MF) level of approximation~\cite{orland1}, and the DPB formalism was reconsidered in bulk liquids at the one-loop order~\cite{orland2}. Different generalized PB approaches based on dielectric continuum but accounting for the multipolar moments of ions were also developed in Refs.~\cite{bohinc1,podgornik}. We have recently extended the DPB formalism of Ref.~\cite{dunyuk} beyond MF level in order to investigate in inhomogeneous electrolytes surface polarization effects on the differential capacity of low dielectric materials~\cite{epl}. This extended DPB (EDPB) approach allowed us to improve the agreement of the PB theory with experimental capacitance data of carbon-based materials in a significant way. However, these generalized approaches that treat the solvent molecules as point-dipoles cannot consider their extended charge structure. As a result, they yield exclusively a local picture of electrostatic interactions in charged systems. At this stage, one should mention the innovative works of Refs.~\cite{Lue1,pincus,bohinc2} that focused on the electrostatics of charges with finite extension at different approximation levels, though the solvent was again considered in these models within the dielectric continuum approach.

In order to overcome these limitations, a microscopic polar liquid model accounting for the discrete charge composition of solvent molecules is needed. In this article, we introduce a microscopic description of non-local electrostatic interactions in polar liquids. We first derive in Sec.~\ref{fiel} the field theoretic model of the polar liquid composed of multipolar solvent molecules of finite size, and embedding polarizable ions modeled as Drude oscillators. Then, we obtain from the saddle point of the model Hamiltonian a non-local PB equation, which is considered in the rest of the article within \textcolor{black}{the linear response regime of polar liquids symmetrically partitioned around weakly charged planar interfaces.} Sec.~\ref{macro} introduces a mapping from the microscopic polar liquid model onto the macroscopic relations of non-local electrostatics. Finally, we investigate in Sec.~\ref{mf} dipolar correlation effects, multipolar contributions to the permittivity of the liquid, and the effect of induced polarizability on the interfacial solvent and salt partition. The limitations of the present theory, its possible extensions and potential applications are thoroughly discussed in the Conclusion part.

\begin{figure}
(a)\includegraphics[width=0.9\linewidth]{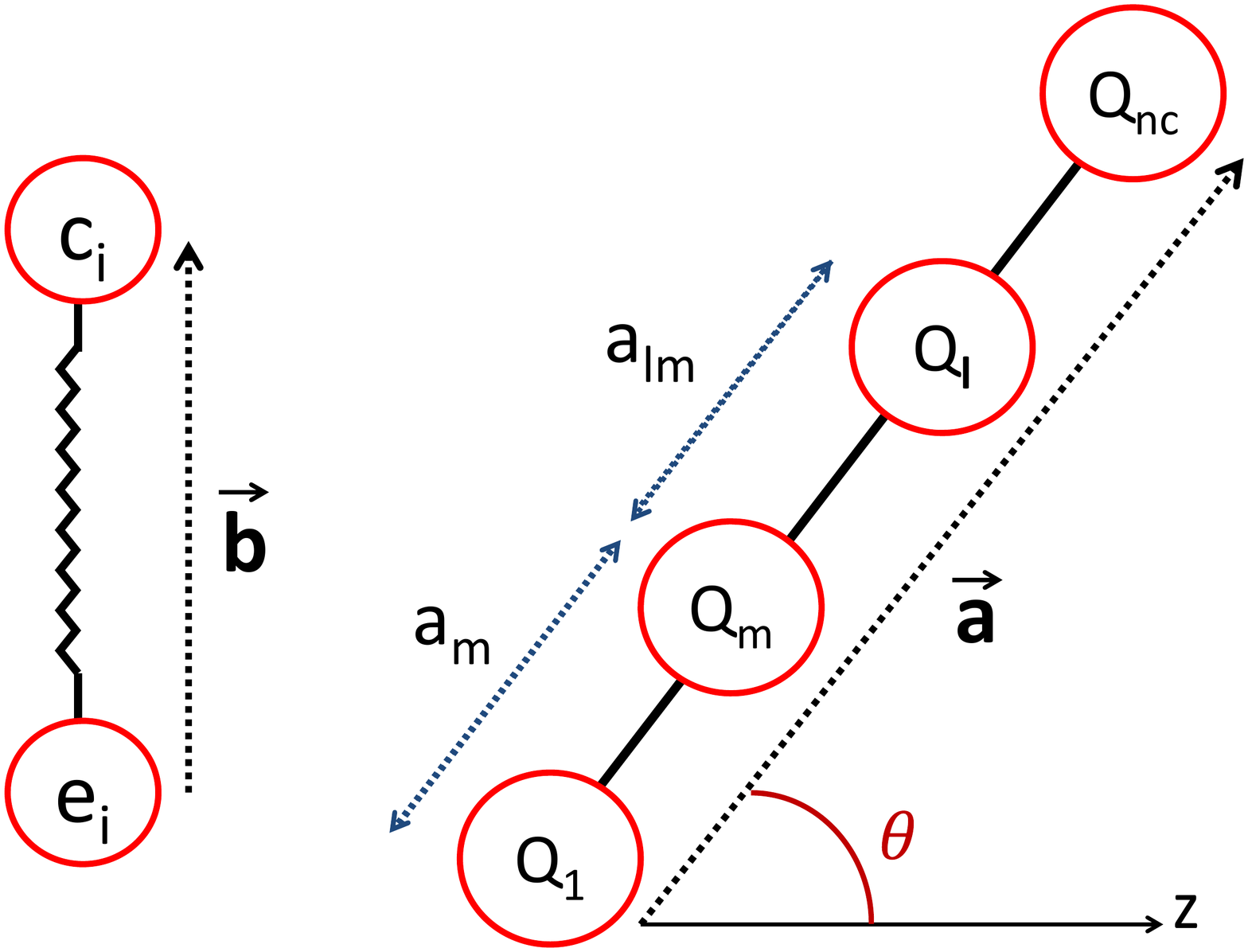}
(b)\includegraphics[width=.9\linewidth]{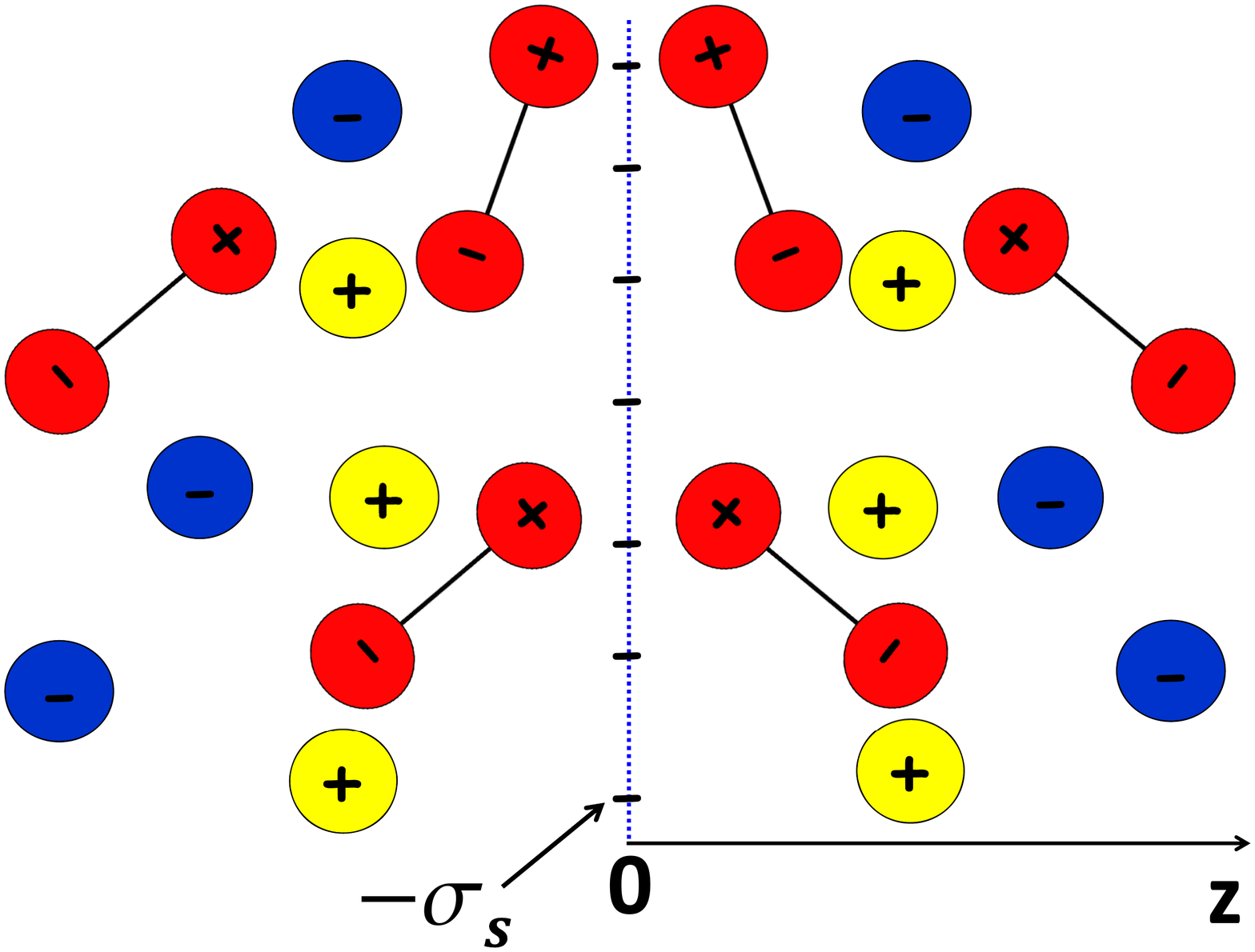}
\caption{(Color online)  (a) Geometry of polarizable ions (left) and multipolar solvent molecules (right). (b) Charged interface.}
\label{fig0}
\end{figure}

\section{Field theoretic model and MF equations}
\label{fiel}

This part is devoted to the derivation of the field theoretic model for a polarizable ion gas of different species, immersed in a solvent composed of polar molecules with a linear charge distribution. A schematic presentation of the solvent charge geometry is given in Fig.~\ref{fig0}(a). Each solvent molecule consists of a rigid rod where $n_c$ elementary charges with valency $Q_l$ are distributed, with the index $l$ running over the elementary charges on the solvent molecule. Our consideration of linear multipolar solvent molecules is motivated by the possibility to incorporate into this geometry the relative dipole/multipole moments of water molecules obtained in quantum molecular calculations~\cite{niu}. This complication will be treated in Sec.~\ref{multi}. We also note that each solvent molecule is overall neutral, that is $\sum_lQ_l=0$.  Furthermore, the distance of the charge $Q_l$ from the first charge is $a_l$, and $a_1=0$ corresponds to the origin of the molecule. Finally, $\ba_{n_c}=\ba=a_{n_c}\bv$ is the vector pointing the end of the molecule, with $a_{n_c}=a$ the total molecular size and the unit vector $\bv$ is parallel to the oriented molecule.

The composition of polarizable ions of $p$ species is also displayed in Fig.~\ref{fig0}(a). Each ion of species $i$ consists of two elementary charges of valency $e_i$ and $c_i$ separated by the distance $\bb$. The electroneutrality condition implies $\sum_i\rho_{ib}(e_i+c_i)=0$. Moreover, the ionic polarizability associated with the deformation of the electronic cloud by the surrounding fields is considered within the Drude oscillator model~\cite{drude},
\be\label{hpol}
h_i\left(\bb\right)=\frac{\bb^2}{4b_{pi}^2},
\ee
where $b_{pi}^2$ is the variance of these oscillations associated with the ions of species $i$. It will be shown in Sec.~\ref{pol} that $b_{pi}^2$ is proportional to the induced polarizability $\alpha$.

The canonical partition function for the system of polar molecules and polarizable ions coupled with electrostatic interactions read
\bea\label{canpart}
Z_c&=&\frac{e^{N_sE_s}}{N_s!\lambda_{Td}^{3N_s}}\int\prod_{k=1}^{N_s}\frac{\mathrm{d}\bo}{4\pi}\mathrm{d}\bx_k\\
&&\times\prod_{i=1}^{p}\prod_{j=1}^{N_i}\frac{e^{N_iE_i}}{N_i!\lambda_{Ti}^{3N_i}}\int\frac{\mathrm{d}\bb_j}{\left(4\pi b_{pi}^2\right)^{3/2}}\mathrm{d}\by_{ij}\;e^{-h_i\left(\bb_j\right)-H(\bu)},\nonumber
\eea
where $N_s$ is the number of solvent molecules,  $N_i$ the number of ions for the species $i$, and $\lambda_{Td}$ and $\lambda_{Ti}$ stand for the thermal wavelength of solvent molecules and ions, respectively. We also introduced the shorthand notation $\bu=\left(\{\bx_k\},\{\ba_k\},\{\by_{ij}\},\{\bb_j\}\right)$ for the configurational vector space, where $\by_{ij}$ and $\bx_k$ stand respectively for the spatial coordinates of the first elementary charge of ions and solvent molecules. Finally, we note that $\bo=(\theta_k,\varphi_k)$ denote the solid angle characterizing the orientation of the $k$th solvent molecule.

The interaction energy is composed of an electrostatic part and a wall contribution, $H(\bu)=H_{el}(\bu)+H_w(\bu)$, where the electrostatic part reads
\be\label{Hel}
H_{el}(\bu)=\frac{1}{2}\int_{\br\br'}\left[\rho_{ic}+\rho_{sc}+\sigma\right]_\br v_c(\br,\br')\left[\rho_{ic}+\rho_{sc}+\sigma\right]_{\br'},
\ee
with $\sigma(\br)$ the fixed charge distribution (see Fig.~\ref{fig0}(b)), and the ionic and solvent charge densities corresponding to the charge compositions in Fig.~\ref{fig0}(a) are respectively defined as
\bea\label{ci}
&&\rho_{ic}(\br)=\sum_{i=1}^p\sum_{j=1}^{N_i}\left[e_i\delta(\br-\by_{ij})+c_i\delta(\br-\by_{ij}-\bb_j)\right]\\
\label{cd}
&&\rho_{sc}(\br)=\sum_{k=1}^{N_s}\sum_{l=1}^{n_s}Q_l\delta(\br-\bx_k-\ba_l).
\eea
Furthermore, the Coulomb potential in Eq.~(\ref{Hel}) is defined as the inverse of the following operator,
\be\label{coulomb}
v_c^{-1}(\br,\br')=-\frac{k_BT}{e^2}\nabla\left[\e(\br)\nabla\delta(\br-\br')\right],
\ee
where $\e(\br)$ is a spatially varying dielectric permittivity. We also note that the bulk Coulomb potential reads $v_c^b(r)=\ell_B/r$, with the Bjerrum length in the air $\ell_B(\br)=e^2/\left[4\pi\e(\br)k_BT\right]\simeq 55$ nm, $e$ the elementary charge, and $T=300$ K the ambient temperature. The inverse of the bulk potential is given by the inverse of the following operator,
\be\label{coulomb2}
{v^b_c}^{-1}(\br,\br')=-\frac{k_BT\eo}{e^2}\Delta\delta(\br-\br').
\ee
In Eq.~(\ref{coulomb2}), $\eo$ stands for the permittivity of the air. We will consider in this work exclusively the case of a uniform background permittivity $\e(\br)=\e_0$, which implies $v_c(\br,\br')=v_c^b(\br-\br')$ and $\ell_B(\br)=\ell_B$. We also note that in Eq.~(\ref{canpart}), we subtracted from the total Hamiltonian the self energy of ions $E_i=\left(e_i^2+c_i^2\right)v_c(0)/2+e_ic_iv_c(b)$  and polar molecules $E_s=Q^2\left[v_c(0)-v_c(a)\right]$ in the air medium.

Finally, the part of the Hamiltonian corresponding to particle-wall interactions read
\be\label{hw1}
H_w(\bu)=\sum_{i=1}^p\sum_{j=1}^{N_i}W_i(\by_{ij},\bb_j)+\sum_{k=1}^{N_s}W_s(\bx_k,\ba_k),
\ee
where we introduced the general wall potentials $W_i(\by_{ij},\bb_j)$ and $W_s(\br,\ba)$ respectively for ions and solvent molecules.  These wall potentials will be used to restrict the phase space accessible to the particles, and also as generatrice functions in order to derive particle number densities. We also note that all energies will be given in units of the thermal energy $k_BT$, the dielectric permittivities in units of the air permittivity $\e_0$, and the surface charges in units of the elementary charge $e$.

In order to simplify the theoretical analysis of the system, one can pass from the density to the field representation by performing a Hubbard-Stratonovich transformation. The grand canonical partition function defined as $Z_G=\sum_{N_s\geq0}\prod_{i=1}^p\sum_{N_i\geq0}e^{\mu_i N_i}e^{\mu_w N_s}Z_c$ takes the form of a functional integral over this fluctuating electrostatic potential, $Z_G=\int \mathcal{D}\phi\;e^{-H[\phi]}$, with the Hamiltonian functional
\bea\label{HamFunc}
&&H[\phi]=\\
&&=\int\mathrm{d}\br\left[\frac{\left[\nabla\phi(\br)\right]^2}{8\pi\ell_B}-i\sigma(\br)\phi(\br)\right]\nonumber\\
&&-\Lambda_s\int\frac{\mathrm{d}\bom}{4\pi}\mathrm{d}\br\;e^{E_s-W_s(\br,\ba)}e^{i\sum_lQ_l\phi(\br+\ba_l)}\nonumber\\
&&-\sum_i\Lambda_i\int\frac{\mathrm{d\bb}}{\left(4\pi b_{pi}^2\right)^{3/2}}\mathrm{d}\br\;e^{-h_i(\bb)+E_i-W_i(\br,\bb)}\nonumber\\
&&\hspace{3.5cm}\times e^{ie_i\phi(\br)+ic_i\phi(\br+\bb)},\nonumber
\eea
where we rescaled the ionic and solvent fugacities as $\Lambda_i=e^{\mu_i}/\lambda_{Ti}^3$ and $\Lambda_s=e^{\mu_s}/\lambda_{Td}^3$.

In order to derive local number densities, we first split the solvent and ion wall potentials into two parts, $W_s(\br,\ba)=W_{s1}(\br)+W_{s2}(\br,\ba)$ and $W_i(\br,\bb)=W_{i1}(\br)+W_{i2}(\br,\bb)$. Passing now from the complex to the real electrostatic potential with the transformation $\phi(\br)\to i\phi(\br)$,  the mean-field level ion and solvent number densities respectively follow by taking the functional derivatives of Eq.~(\ref{HamFunc}) with respect to $W_{i1}(\br)$ and $W_{s1}(\br)$,
\bea\label{ipar}
\rho_{ip}(\br)&=&\Lambda_i\int\frac{\mathrm{d\bb}}{\left(4\pi b_{pi}^2\right)^{3/2}}\;e^{-h_i(\bb)+E_i-W_i(\br,\bb)}\\
&&\hspace{2.7cm}\times e^{-e_i\phi(\br)-c_i\phi(\br+\bb)}\nonumber\\
\label{solpar}
\rho_{sp}(\br)&=&\Lambda_s\int\frac{\mathrm{d}\bom}{4\pi}\;e^{E_s-W_s(\br,\ba)}e^{-\sum_lQ_l\phi(\br+\ba_l)}.
\eea
In terms of the same real electrostatic potential, the MF-level saddle point equation $\delta H[\phi]/\delta \phi(\br)=0$ takes the form of a generalized PB equation,
\be\label{eq1}
\Delta\phi(\br)+4\pi\ell_B\left[\sigma(\br)+\sum_i\rho_{ic}(\br)+\rho_{sc}(\br)\right]=0,
\ee
where we introduced the ionic and solvent charge densities
\bea\label{ionch}
\rho_{ic}(\br)&=&\rho_{ib}\int\frac{\mathrm{d}\bb}{\left(4\pi
b_{pi}^2\right)^{3/2}}\;e^{-h_i(\bb)}\\
&&\hspace{2.0cm}\times\left[e_ie^{-W_i(\br,\bb)}e^{-e_i\phi(\br)-c_i\phi(\br+\bb)}\right.\nonumber\\
&&\left.\hspace{1.3cm}+c_ie^{-W_i(\br-\bb,\bb)}e^{-e_i\phi(\br-\bb)-c_i\phi(\br)}\right]\nonumber\\
\label{solch}
\rho_{sc}(\br)&=&\rho_{sb}\int\frac{\mathrm{d}\bom}{4\pi}\sum_mQ_me^{-W_s(\br-\ba_m,\bom)}\\
&&\hspace{2.45cm}\times e^{-\sum_lQ_l\phi(\br+\ba_l-\ba_m)}\nonumber,
\eea
and used the MF relations between the charge densities and fugacities $\Lambda_ie^{E_i}=\rho_{ib}$ and $\Lambda_se^{E_s}=\rho_{sb}$. These relations follow from the bulk limits of Eqs.~(\ref{ipar}) and~(\ref{solpar}). We also note that in this work, the particle and charge partition functions for ions and solvent molecules will be related to the local densities in Eqs.~(\ref{ipar})-(\ref{solpar}) and Eqs.~(\ref{ionch})-(\ref{solch}) according to $k(\br)=\rho(\br)/\rho_{b}$.

\begin{figure}
\includegraphics[width=1.15\linewidth]{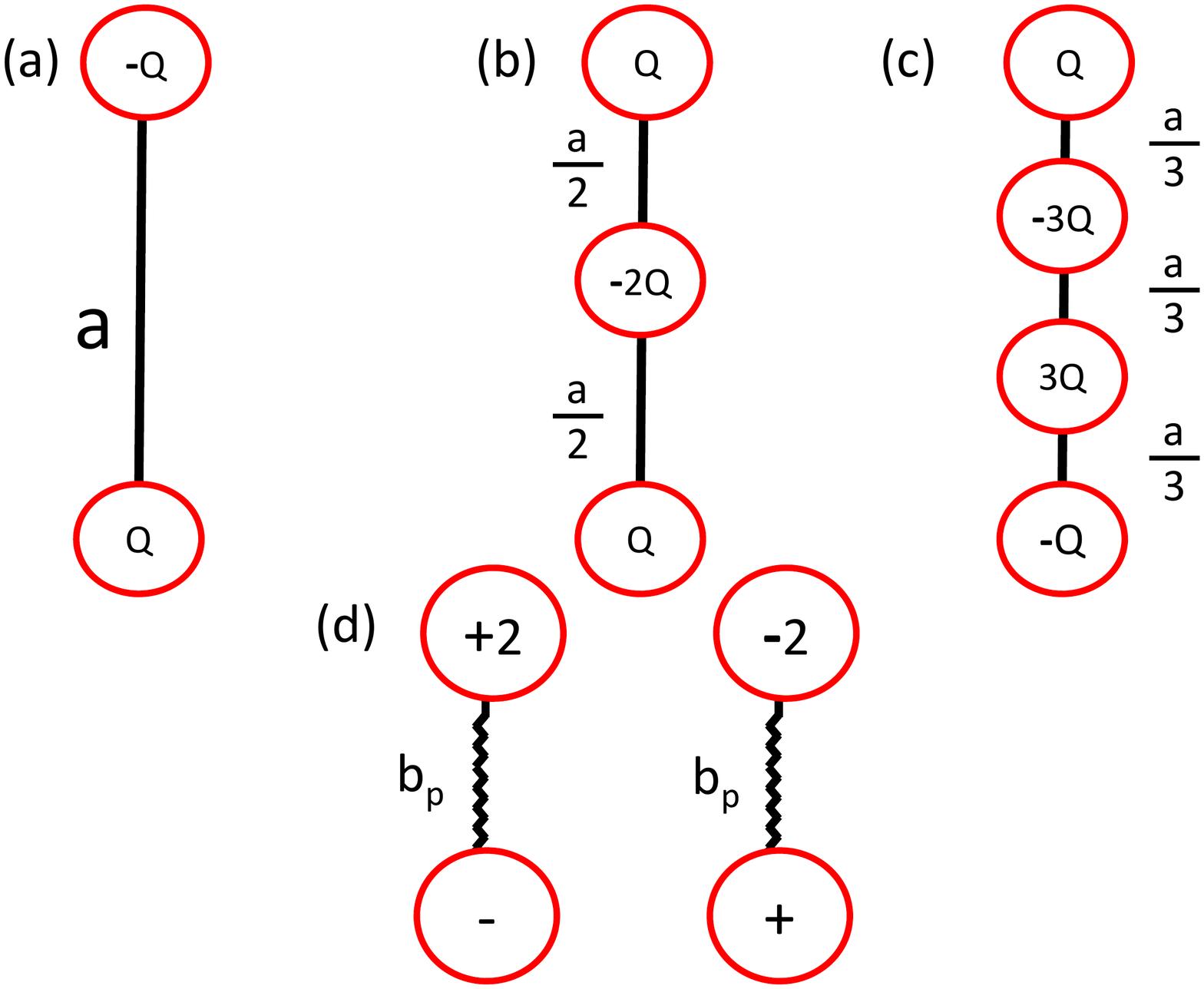}
\caption{(Color online) Charge composition of (a) linear dipoles, (b) quadrupoles , (c) octupoles, and (d) polarizable ions considered in Sec.~\ref{pol}.}
\label{fig6}
\end{figure}

The dependence of the solvent charge densities in Eq.~(\ref{solch}) on the values of the electrostatic potential at different points around $z$ makes Eq.~(\ref{eq1}) a non-local Poisson-Boltzmann (NLPB) equation that embodies the non-local dielectric response of the polar liquid at the molecular level of precision. \textcolor{black}{We also note that for ions with vanishing polarizability ($b_p=0$) and solvent molecules of dipolar geometry (see Fig.~\ref{fig6}(a)), by expanding the argument of the potential in the exponential of Eq.~(\ref{solch}) at the order $O(a^2$), the NLPB equation~(\ref{eq1}) tends to the DPB equation derived in Ref.~\cite{dunyuk}.}

We will investigate in this article the MF theory of non-local electrostatic interactions for polar liquids in contact with a charged planar interface located at $z=0$, and corresponding to a surface charge distribution $\sigma(\br)=-\sigma_s\delta(z)$ with $\sigma_s>0$.  The negatively charged wall splits the space accessible to the liquid into two regions $z<0$ and $z>0$, with equal dipolar and ionic bulk concentrations on each side (see Fig.~\ref{fig0}(b)). The rotational restriction for dipoles will be considered exclusively at the end of the Sec.~\ref{efper}, and we will assume that the interface at $z=0$ is penetrable in the rest of the article, that is $W_s(\br,\bom)=W_i(\br)=0$.  We also note that due to the translational symmetry within the $(x,y)$ plan, the electrostatic potential depends only on the separation from the wall at $z=0$, i.e. $\phi(\br)=\phi(z)$.

Moreover, we will consider exclusively \textcolor{black}{the linear response regime corresponding to weak surface charges}. \textcolor{black}{By expanding Eq.~(\ref{eq1}) at the linear order in the electrostatic potential $\phi(z)$, and passing from the azimuthal angle $\theta$ to the projection of the dipole orientation on the $z-$axis with the transformation $a_z=a\cos\theta$, one obtains the following non-local differential equation,}
\bea\label{eql2}
&&\Delta\phi_0(z)+4\pi\ell_B\sigma(z)-\e_w\kappa_i^2\phi_0(z)-\kappa_s^2\phi(z)\\
&&-4\pi\ell_B\rho_{sb}\sum_{l\neq m}Q_lQ_m\int_{-a_{lm}}^{a_{lm}}\frac{\mathrm{d}a_z}{2a_{lm}}\phi_0(z+a_z)\nonumber\\
&&-4\pi\ell_B\sum_i\rho_{ib}e_ic_i\int_{-\infty}^{+\infty}\frac{\mathrm{d}b_z}{\sqrt{4\pi b_{pi}^2}}e^{-h_i\left(b_z\right)}\nonumber\\
&&\hspace{2.8cm}\times\left[\phi_0(z+b_z)+\phi_0(z-b_z)-2\phi_0(z)\right]\nonumber\\
&&=0,\nonumber
\eea
where $\kappa_i^2=4\pi\ell_B\sum_i\rho_{ib}q_i^2/\e_w$ is the ionic screening parameter, with $q_i=e_i+c_i$ the total charge of each ion of species $i$, and $\kappa_s^2=4\pi\ell_B\rho_{sb}\sum_lQ_l^2$ the screening parameter associated with solvent charges. We introduced above the notation $a_{lm}=a_l-a_m$ for the separation distance between the charges $l$ and $m$ on the solvent molecule (see Fig.~\ref{fig0}), and also the dielectric permittivity in the bulk solvent medium $\e_w$  that will be defined in Sec.~\ref{macro} (see Eq.~(\ref{ew})). The Bjerrum length in the water is indeed related to the one in the air medium as $\ell_w=\ell_B/\e_w$. Furthermore, the naught in $\phi_0(z)$ means that deriving Eq.~(\ref{eql2}), we neglected the rotational penalty for the solvent molecules  in the region $z<a$. We finaly note that all numerical results will be obtained for monovalent ions $q_i=1$ (until Sec.~\ref{pol} on polarizable ions), and the model parameters of the dipolar solvent molecules will be chosen as $Q=1$ and $a=1$ {\AA}, unless otherwise stated.

\section{Mapping to the macroscopic formulation of non-local electrostatics}
\label{macro}

We introduce in this part a mapping from the microscopic model of Eq.~(\ref{eql2}) onto the macroscopic formulation of non-local electrostatics. This mapping will allow us to relate the effective permittivity of the polar medium to the polarization charges in the liquid. By defining first the kernel operator
\bea\label{eq5}
G^{-1}(z,z')&=&\frac{-\partial_z^2+\e_w\kappa_i^2}{4\pi\ell_B}\delta(z-z')+\frac{\kappa_s^2}{4\pi\ell_B}\delta(z-z')\\
&&+\rho_{sb}\sum_{l\neq m}Q_lQ_m\int_{-a_{lm}}^{a_{lm}}\frac{\mathrm{d}a_z}{2a_{lm}}\delta(z'-z-a_z)\nonumber\\
&&+\sum_i\rho_{ib}e_ic_i\int_{-\infty}^{+\infty}\frac{\mathrm{d}b_z}{\sqrt{4\pi b_{pi}^2}}e^{-h_i\left(b_z\right)}\nonumber\\
&&\hspace{0cm}\times\left[\delta(z'-z-b_z)+\delta(z'-z+b_z)-2\delta(z'-z)\right],\nonumber
\eea
and using the definition of the Green's function
\be\label{eq6}
\int\mathrm{dz'}G^{-1}(z,z')G(z',z'')=\delta(z-z''),
\ee
the linear NLPB equation~(\ref{eql2}) that can be reexpressed as
\be\label{eq7}
\int_{-\infty}^{\infty}\mathrm{d}z'G^{-1}(z,z')\phi_0(z')=\sigma(z')
\ee
can be inverted, and the solution expressed in the form
\be\label{eq8}
\phi_0(z)=\int_{-\infty}^{+\infty}\mathrm{d}z'G(z,z')\sigma_s(z').
\ee
Using now the relations~(\ref{eq5}) and~(\ref{eq6}), the Green's function can be derived in 1D Fourier space as
\be\label{eq10}
G(z-z')=4\ell_w\int_0^\infty\mathrm{d}k\frac{\cos\left[k(z-z')\right]}{\kappa_i^2+k^2\te(k)/\e_w},
\ee
where we introduced the dielectric permittivity function in Fourier space,
\bea\label{eq11}
\te(k)&=&1+\frac{\kappa_s^2}{k^2}+\frac{4\pi\ell_B\rho_{sb}}{k^2}\sum_{l\neq m}Q_lQ_m\frac{\sin\left(ka_{lm}\right)}{ka_{lm}}\\
&&+\frac{8\pi\ell_B}{k^2}\sum_i\rho_{ib}e_ic_i\left(e^{-b_{pi}^2k^2}-1\right).\nonumber
\eea
We note that the second-third and the forth terms on the r.h.s. of Eq.~(\ref{eq11}) correspond respectively to the charge structure factor of solvent molecules and polarizable ions in Fourier space. The bulk permittivity introduced in Eq.~(\ref{eql2}) is precisely defined as the infrared (IR) limit of Eq.~(\ref{eq11}), $\e_w\equiv\te(k\to0)$, and it is given by
\be\label{ew}
\e_w=1-\frac{2\pi\ell_B\rho_{sb}}{3}\sum_{l\neq m}Q_lQ_ma_{lm}^2-8\pi\ell_B\sum_i\rho_{ib}b_{pi}^2e_ic_i.
\ee

Plugging now the Green's function~(\ref{eq10}) into Eq.~(\ref{eq8}), the electrostatic potential follows in the form
\be\label{eq12}
\phi_0(z)=-\frac{2}{\pi\mu_i q_i}\int_0^\infty\mathrm{d}k\frac{\cos(kz)}{\kappa_i^2+k^2\te(k)/\e_w},
\ee
where $\mu_i=1/(2\pi q_i\ell_w\sigma_s)$ stands for the ionic Gouy-Chapman length, and the net electrostatic field $E(z)=\phi'_0(z)$ reads
\be\label{eq13}
E(z)=\frac{2}{\pi\mu_i q_i}\int_0^\infty\mathrm{d}k\frac{k\sin(kz)}{\kappa_i^2+k^2\te(k)/\e_w}.
\ee
\textcolor{black}{We note that the potential in Eq.~(\ref{eq12}) characterized by a diffuse permittivity function is similar in form to Eq. (3.16) of Ref.~\cite{Kor2} where a phenomenological dielectric permittivity function was used in order to investigate non-local electrostatic effects on the charge storage ability of metallic capacitors.}

Using Eqs.~(\ref{eq6}) and~(\ref{eq10}), one can show that Eq.~(\ref{eq5}) can be recasted in the form of a non-local electrostatic kernel
\be\label{eq14}
G^{-1}(z,z')=\frac{1}{4\pi\ell_B}\left[-\partial_z\e(z-z')\partial_{z'}+\e_w\kappa_i^2\delta(z-z')\right],
\ee
where the non-local dielectric permittivity function is simply the inverse Fourier transform of Eq.~(\ref{eq11}),
\be\label{eq15}
\e(z-z')=\delta(z-z')+4\pi\ell_B\chi(z-z'),
\ee
with susceptibility function
\bea\label{eq16}
\chi(z)&=&-\frac{\kappa_s^2}{8\pi\ell_B}|z|\\
&&+\rho_{sb}\sum_{l\neq m}\frac{Q_lQ_m}{8a_{lm}}\left\{(z-a_{lm})^2\mathrm{sign}(z-a_{lm})\right.\nonumber\\
&&\hspace{2cm}-\left.(z+a_{lm})^2\mathrm{sign}(z+a_{lm})\right\}\nonumber\\
&&+2\sum_i\rho_{ib}b_{pi}e_ic_i\left\{\frac{|z|}{2b_{pi}}\mathrm{Erfc}\left(\frac{|z|}{2b_{pi}}\right)-\frac{h_i(z)}{\sqrt\pi}\right\}\nonumber,
\eea
where the distortion energy of polarizable ions $h_i(z)$ is given by Eq.~(\ref{hpol}). The equation~(\ref{eq16}) is the key result relating the microscopic polar liquid model of Eq.~(\ref{HamFunc}) to the macroscopic formulation of non-local electrostatics. A simpler form for this susceptibility function will be given for the case of simple dipolar liquids with point ions in Sec.~\ref{col} and for polarizable ions embedeed in the dipole liquid in Sec.~\ref{pol} .

We now note that in terms of the non-local dielectric displacement field
\be\label{eq17}
D(z)=\int_{-\infty}^{\infty}\mathrm{d}z'\e(z-z')E(z'),
\ee
the equation~(\ref{eq7}) can be rewritten as
\be\label{eq18}
-\partial_zD(z)=4\pi\ell_B\sigma(z)-\e_w\kappa_i^2\phi_0(z).
\ee
Neglecting the ionic screening term in the region $\kappa_iz\ll1$ and integrating Eq.~(\ref{eq18}), one gets for the induction field
\be\label{eq19}
D(z)=2\pi\ell_B\sigma_s\mathrm{sign}(z).
\ee
We now define the polarization field $P(z)$ through the usual non-local dielectric response equation
\be\label{eq20}
P(z)=\int_{-\infty}^{\infty}\mathrm{d}z'\chi(z-z')E(z').
\ee
In the next section where we investigate the simplest case of ions without polarizability in a dipole liquid, it will be explicitly shown that the function $\chi(z-z')$ brings the contribution from individual dipoles to the total dielectric correlation function, with a characteristic correlation length of the same order as the molecular size. Furthermore, by using Eqs.~(\ref{eq15}) and~(\ref{eq19}), the polarization field introduced in Eq.~(\ref{eq20}) can be related to the displacement and total electrostatic fields as
\be\label{eq21}
P(z)=\frac{D(z)-E(z)}{4\pi\ell_B}.
\ee
Reexpressed in the form $E(z)=D(z)-4\pi\ell_BP(z)$, this well-known macroscopic relation can be interpreted as follows. In a polar medium in contact with a fixed charge source (e.g. the surface charge located at $z=0$), the local field experienced by a test ion at the position $z$ is the superposition of the induction field $D(z)$ generated by the surface charge (i.e. the field in the air medium), and the reaction field induced by the polarizable molecules in response to this induction field. The reduction of the latter by the polarization field is the so-called \textit{dielectric screening} effect.

Furthermore, we note that in the same region $\kappa_iz\ll1$, the electrostatic field in Eq.~(\ref{eq13}) becomes
\be\label{eq22}
E(z)\simeq\frac{\ew}{q_i\mu_i\e_{\mathrm{eff}}(z)},
\ee
where we introduced the local effective dielectric permittivity function
\be\label{eq23}
\e_{\mathrm{eff}}(z)=\frac{\pi}{2}\left/\int_0^\infty\frac{\mathrm{d}k}{k}\frac{\sin(kz)}{\te(k)}.\right.
\ee
Injecting now the relations~(\ref{eq19}) and~(\ref{eq22}) into Eq.~(\ref{eq21}), one obtains a relation between the macroscopic dielectric permittivity function and the polarization field,
\be\label{eq24}
P(z)=\frac{\sigma_s}{2}\left[\mathrm{sign}(z)-\frac{1}{\e_{\mathrm{eff}}(z)}\right].
\ee
Moreover, by substituting the relations~(\ref{eq20}) and~(\ref{eq21}) into Eq.~(\ref{eq18}), and using the non-local PB equation~(\ref{eql2}), one obtains an expression relating the variations of the polarization field to the polarization densities,
\be\label{eq25}
\frac{\partial P(z)}{\partial z}=\rho_{sb}k_{sc}(z)+\sum_i\rho_{ib}k_{pc}^{(i)}(z),
\ee
where the solvent charge partition function Eq.~(\ref{solch}) takes in the linear potential approximation the form
\bea\label{eq26}
k_{sc}(z)&=&-\sum_lQ_l^2\phi_0(z)\\
&&-\sum_{l\neq m}Q_lQ_m\int_{-a_{lm}}^{a_{lm}}\frac{\mathrm{d}a_z}{2a_{lm}}\phi_0(z+a_z),\nonumber
\eea
and we also introduced the part of the ion charge partition function  associated with the ionic polarizability (i.e. the sixth term on the lhs of Eq.~(\ref{eql2})),
\bea\label{eq261}
k_{pc}^{(i)}(z)&=&-e_ic_i\int_{-\infty}^{+\infty}\frac{\mathrm{d}b_z}{\sqrt{4\pi b_{pi}^2}}e^{-h_i\left(b_z\right)}\\
&&\hspace{1.3cm}\times\left[\phi_0(z+b_z)+\phi_0(z-b_z)-2\phi_0(z)\right].\nonumber
\eea
We now note that the inverse permittivity function that allows us to invert Eq.~(\ref{eq17}) as
\be\label{eq262}
E(z)=\int_{-\infty}^{\infty}\mathrm{d}z'\e^{-1}(z-z')D(z'),
\ee
is given by
\be\label{eq263}
\e^{-1}(z)=\frac{1}{\pi}\int_0^\infty\frac{\mathrm{d}k\cos(kz)}{1+4\pi\ell_B\tchi(k)},
\ee
where the Fourier transformed susceptibility is defined through the relations~(\ref{eq11}) and~(\ref{eq15}) as $\tchi(k)=\left[\te(k)-1\right]/(4\pi\ell_B)$.

Thorough the expression~(\ref{eq262}), one can see the inverse dielectric permittivity $\e^{-1}(z-z')$ as the dielectric correlation function containing the whole information on the polarizability of the solvent. Comparing the inverse permittivity in Eq.~(\ref{eq263}) with the effective permittivity Eq.~(\ref{eq23}), we find that both functions are related as $\e^{-1}(z)=\partial_z\left[2\e_\mathrm{eff}(z)\right]^{-1}$. With the use of Eqs.~(\ref{eq24}) and~(\ref{eq25}), this relation shows that the inverse permittivity is related to the normalized charge densities associated with the polarizable molecules by the simple equation
\bea\label{eq2632}
\e^{-1}(z)=\delta(z)-\frac{1}{\sigma_s}\left[\rho_{sb}k_{sc}(z)+\sum_i\rho_{ib}k_{pc}^{(i)}(z)\right].
\eea
Injecting this relation into Eq.~(\ref{eq262}), we obtain a relation that expresses the modification of the induction field by the polar liquid as the convolution of the former with the charge density of polar molecules over the whole space,
\bea\label{eq264}
E(z)&=&D(z)-\frac{1}{\sigma_s}\int_{-\infty}^{\infty}\mathrm{d}z'\left[\rho_{sb}k_{sc}(z-z')\right.\\
&&\hspace{3cm}\left.+\sum_i\rho_{ib}k_{pc}^{(i)}(z-z')\right]D(z').\nonumber
\eea
By considering now the explicit form of the induction field~(\ref{eq19}) in Eq.~(\ref{eq264}), one finally gets with the use of Eq.~(\ref{eq22}) a relation between the local value of the dielectric permittivity $\e_\mathrm{eff}(z)$ and the integrated polarization density,
\be\label{eq27}
\frac{1}{\e_{\mathrm{eff}}(z)}=1-\frac{2}{\sigma_s}\int_0^z\mathrm{d}z'\left[\rho_{sb}k_{sc}(z')+\sum_i\rho_{si}k_{pc}^{(i)}(z')\right]
\ee
for $z\geq0$. Deriving Eq.~(\ref{eq27}), we used the reflection symmetry of the densities with respect to the interface, i.e. $k_{sc}(-z)=k_{sc}(z)$ and  $k_{pc}^{(i)}(-z)=k_{pc}^{(i)}(z)$.

According to the relation~(\ref{eq27}), in the linear response regime, local deviations of the dielectric permittivity from the permittivity of the air are  related to the accumulated polarization charge of solvent molecules and ions between the considered point in the liquid and the charged plan. As a result, the effective permittivity tends on the surface to the permittivity of the air, that is, the immediate vicinity of the interfacial area is characterized by a dielectric void. The manifestation of this peculiarity absent in local electrostatic theories~\cite{dunyuk,orland1,epl} but present in MD simulations~\cite{prlnetz,langnetz,Hans} and AMF experiments~\cite{expdiel} indicates that the proper consideration of the extended charge structure of solvent molecules in our model is the key ingredient to recover the correct dielectric response of the water solvent. This point will be elaborated in further detail in the following parts.

\section{Non-local electrostatic effects in dipolar liquids at charged interfaces}
\label{mf}

We investigate in this part non-local electrostatic effects on the dielectric response of the liquid, and the solvent and salt partitions at the charged interface. The geometry of the dipolar solvent molecules composed of two point-charges with valency $\pm Q$ is presented in Fig.~\ref{fig6}(a). For this specific solvent charge geometry and non-polarizable ions (i.e. $b_{ip}=0$), we will first enlighten in Sec.~\ref{col} a collective dielectric response mechanism at the scale of single dipoles. Then in Sec.~\ref{sd}, we will show that the same mechanism is responsible for the formation of successive hydration layers around the charged surface, which will be shown in Sec.~\ref{efper} to explain the characteristic shape of the transverse permittivity profiles observed in MD simulations~\cite{prlnetz,langnetz,Hans}. At the next step in Sec.~\ref{multi}, we will estimate the contribution of the multipolar moments of water molecules to the dielectric permittivity of the liquid. Finally, the effect of induced ion polarizability on the salt partition at charged interfaces will be discussed in Sec.~\ref{pol}.

\subsection{Collective dielectric response mechanism}
\label{col}

We examine in this part the non-local dielectric screening mechanism discussed in Sec.~\ref{macro} in terms of individual dipole interactions. To this end, we note that for the dipolar charge composition in Fig.~\ref{fig6}(a), the dielectric susceptibility function defined in Eq.~(\ref{eq16}) is given by $\chi(z-z')=2p_0^2\rho_{sb}/aC(z-z')$, with the dipole moment $p_0=Qa$ and the adimensional susceptibility
\be\label{eq29}
C_1(z-z')=\frac{1}{4}\left(1-\frac{|z-z'|}{a}\right)^2\theta\left(a-|z-z'|\right).
\ee
The susceptibility function~(\ref{eq29}) shows that the polarizability of the solvent liquid takes place over a finite region determined by the solvent molecular size.

We now note that by expanding the denominator of the integrand of Eq.~(\ref{eq263}) in powers of $\kappa_sa$, and using the convolution theorem in Fourier space, the inverse permittivity function can be rewritten in the form of a geometric series,
\bea\label{eq292}
\e^{-1}(z-z')=\delta(z-z')+\frac{1}{a}\sum_{n\geq 1}(-1)^n\left(\kappa_sa\right)^{2n}C_n(z-z'),\nonumber\\
\eea
where the high order correlation functions for $n>1$ are related with the susceptibility in Eq.~(\ref{eq29}) by the recurrence relation
\be\label{recC}
C_n(z-z')=\int_{-\infty}^\infty\frac{\mathrm{d}z''}{a}C_1(z-z'')C_{n-1}(z''-z').
\ee
One notices that the first term on the rhs of Eq.~(\ref{eq292}) corresponds to local dielectric correlations, and the second term of order $O\left((\kappa_sa)^2\right)$ that introduces the non-local dielectric response extends the range of these correlations by one molecular size $a$. It is also shown in Appendix~\ref{dil} that at the same order $O\left((\kappa_sa)^2\right)$, the function  $C(z)$ is proportional to the charge density of a single dipole interacting with the charged surface in the air medium, i.e. $k_{sc}(z)=4Qa/\mu_sC(z)$, where we introduced the dipolar Gouy-Chapman length
\be\label{eq293}
\mu_s=\left(2\pi Q\ell_B\sigma_s\right)^{-1}.
\ee
Thus, the non-local contribution to the dielectric correlation function in Eq.~(\ref{eq292}) results in the dilute solvent regime from the response of individual dipoles to the induction field. Moreover, from the recurrence relation~(\ref{recC}), one finds that the next contribution of order $O\left((\kappa_sa)^4\right)$ is characterized by a total range of $2a$,
\bea\label{c2}
C_2(z)&=&\frac{1}{480}\left\{\left(-\bz^5-10\bz^4+40\bz^3-40\bz^2+12\right)\theta(1-\bz)\right.\nonumber\\
&&\hspace{1cm}\left.+\left(2-\bz\right)^5\theta(\bz-1)\theta(2-\bz)\right\},
\eea
with the rescaled distance $\bz=z/a$. To conclude, each correction term of this expansion corresponding to a higher order contribution in the solvent concentration extends the range of non-local polarization effects by one molecular size.
\begin{figure}
\includegraphics[width=0.9\linewidth]{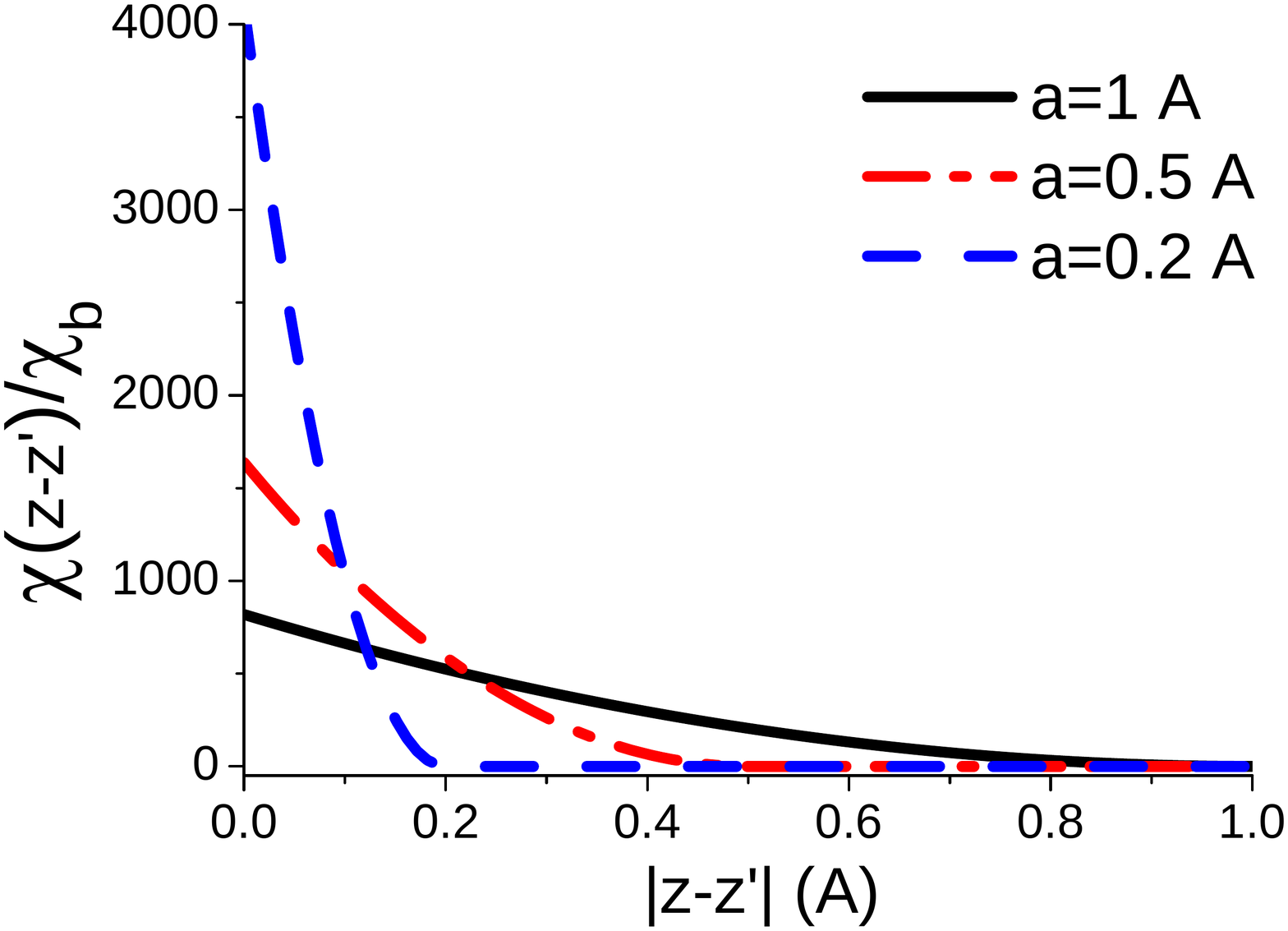}
\caption{(Color online) Dielectric susceptibility function obtained from Eq.~(\ref{eq29}) normalized by the susceptibility of the dielectric continuum approach $\chi_b=\rho_{sb}p_0^2/3$ for $p_0=Qa=1$ {\AA} kept constant, and different values of $a$.}
\label{fig1}
\end{figure}

Substituting now the relation~(\ref{eq292}) into Eq.~(\ref{eq262}), one gets for the electrostatic field the following expansion in powers of the solvent density,
\bea\label{eq294}
E(z)&=&D(z)-4\pi\ell_B\sum_{n\geq 1}\left(\kappa_sa\right)^{2n}P_n(z),
\eea
where the lowest contribution to the dielectric response is given by the dimensionless polarization field $P_1(z)=\int\mathrm{d}z'C_1(z-z')D(z')/(4\pi\ell_Ba)$, and the higher order terms with $n>1$ are obtained from the relation
\be\label{recP}
P_n(z)=-\int_{-\infty}^\infty\frac{\mathrm{d}z'}{a}C_1(z-z')P_{n-1}(z').
\ee
Considering our preceding discussion, the expansion in Eq.~(\ref{eq294}) indicates that at the lowest order in the solvent density $O\left((\kappa_sa)^2\right)$, the dielectric screening is induced by the response of individual dipoles to the induction field .

Furthermore, from the recurrence relation~(\ref{recP}), it follows that the next correction of order $O\left((\kappa_sa)^4\right)$ in Eq.~(\ref{eq294}) with an alternating sign is induced in turn by the response of all individual dipoles to the polarization field of order $O\left((\kappa_sa)^2\right)$, and this correction to the polarization field positively adds to the induction field. Thus, collective effects come into play at the order $O\left((\kappa_sa)^4\right)$. Indeed, the expansion~(\ref{eq294}) shows that the same self consistent relation between the response of individual dipoles and the resulting modification of the polarization field can be extrapolated to higher orders $n$. We also note the apparition of this cooperative behavior at the MF level of approximation is rather remarkable.

We finally show in Fig.~\ref{fig1} that while keeping the dipole moment $p_0$ constant and decreasing the dipole size $a$, the susceptibility function becomes gradually more localized. Taking the dielectric continuum limit $a\to0$, the non-locality disappears, and one gets from Eq.~(\ref{eq29})
\be\label{eq30}
\lim_{a\to0}\chi(z-z')=\chi_b\delta(z-z'),
\ee
with $\chi_b=p_0^2\rho_{sb}/3$ the dielectric susceptibility of the PB formalism in the region $\kappa_iz\ll 1$. Thus, in the point dipole limit, the dielectric response of the polar medium to an external field becomes local, and one recovers the dielectric continuum result $E(z)=\e_wD(z)$ for the net external field.

\subsection{Interfacial solvent densities}
\label{sd}

For \textcolor{black}{low surface charges corresponding to weak potentials $\phi(z)<1$}, the excess number density that we define as $\delta k_{sp}(z)=k_{sp}(z)-1$ follows from Eq.~(\ref{solpar}) in the form
\be\label{eq32}
\delta k_{sp}(z)=\frac{2}{\pi\mu_s}\int_0^\infty\frac{\mathrm{d}k\cos(kz)}{\e_w\kappa_i^2+k^2\te(k)}\left[1-\frac{\sin(ka)}{ka}\right],
\ee
where the function $\te(k)$ introduced in Eq.~(\ref{eq11}) is given by
\be\label{eq28}
\te(k)=1+\frac{\kappa_s^2}{k^2}F(ka),
\ee
with the dipolar screening parameter $\kappa_s^2=8\pi\ell_B\rho_{sb}Q^2$ and charge structure factor $F(x)=1-\sin(x)/x$. We  note that for the dipolar geometry and at the physiological solvent concentration $\rho_{sb}=55$ M, the bulk dielectric permittivity of Eq.~(\ref{ew}) is given by the Debye-Langevin equation $\e_w=1+4\pi\ell_BQ^2a^2\rho_{sb}/3=76.75$.

\begin{figure}
\includegraphics[width=1.05\linewidth]{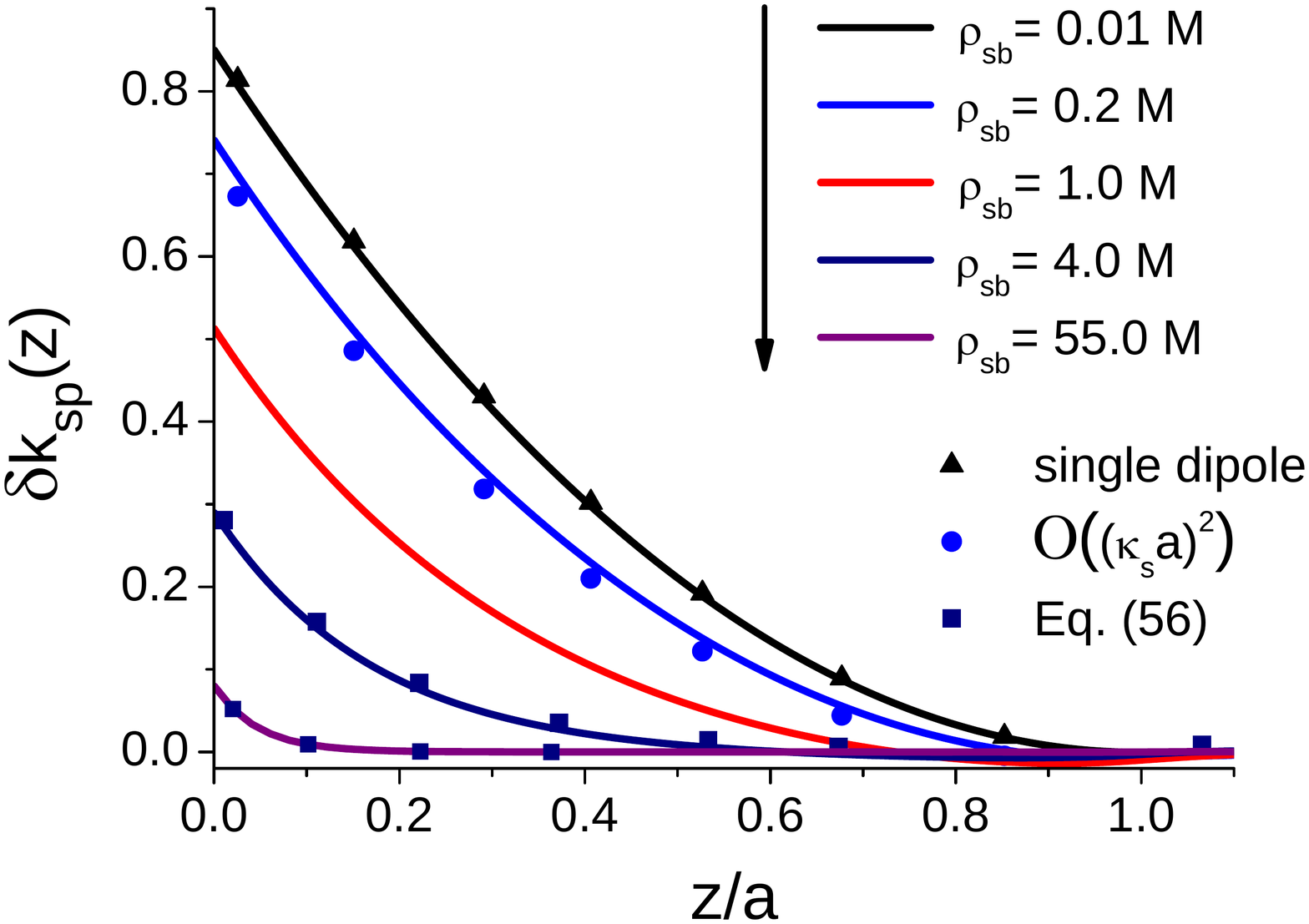}
\caption{(Color online) Solid curves : excess solvent partition function Eq.~(\ref{eq32}) against the rescaled distance from the charged interface at vanishing salt density and several bulk concentrations. The surface charge is $\sigma_s=0.05$ $\mbox{e nm}^{-2}$. Triangles denote the one-particle density profile of Eq.~(\ref{eq34}), circles are from Eq.~(\ref{eq35}) at the order $O(\left(\kappa_sa)^2\right)$, and squares mark the high concentration limit in Eq.~(\ref{eq37}).}
\label{fig2}
\end{figure}

The excess solvent partition function in Eq.~(\ref{eq32}) is displayed in Fig.~\ref{fig2} for various dipolar bulk concentrations. First of all, it is seen that at all concentrations $\rho_{sb}$, there exists a net solvent excess in the neighborhood of the interface, and the dipolar attraction is weakened with an increase of $\rho_{sb}$. Then, we notice that beyond the dilute solvent regime $\rho_{sb}\gtrsim 1$ M, the increase of the bulk solvent density also results in a reduction of the range of the interfacial dipolar attraction. In order to elucidate these points, we will consider the opposite limits of dilute and concentrated solvents where close form expressions for solvent densities can be derived.

In the dilute solvent regime $\kappa_sa<1$, by expanding in the region $\kappa_iz\ll1$ the integrand of Eq.~(\ref{eq32}) in powers of $\kappa_sa$, or comparing Eqs.~(\ref{eq2632}) and~(\ref{eq292}) with the equality between the charge and excess number partition functions
\be\label{cp}
k_{sc}(z)=2Q\delta k_{sp}(z),
\ee
one obtains for the excess solvent partition function
\be\label{eq35}
\delta k_{sp}(z)=\frac{2a}{\mu_s}\sum_{n\geq1}(-1)^{n-1}\left(\kappa_sa\right)^{2n-2}C_n(z),
\ee
where the functions $C_n(z)$ are introduced in Eq.~(\ref{recC}). First of all, we recognize in the first term of the series the one particle dipolar partition function Eq.~(\ref{eq34}) derived in Appendix~\ref{dil}. This contribution was shown in the previous part to result in the polarization field of individual dipoles (the first term of the series in Eq.~(\ref{eq294})). This limiting law is reported in Fig.~\ref{fig2} for $\rho_{sb}=0.01$ M. Thus in the dilute solvent regime, the interfacial dipolar excess extends exactly over the distance $a$, and the dipole behaves as an overall neutral molecule for $z>a$.

Then, noting that the function $C_2(\bar z)$ of Eq.~(\ref{c2}) is positive for all $\bar z$, one sees that the next leading term in Eq.~(\ref{eq35}) of order $O\left((\kappa_sa)^2\right)$ that becomes relevant for $\kappa_sa\sim1$ (or $\rho_{sb}\sim 0.1$ M for $a=1$ {\AA}) corresponds to a net reduction of the single dipole partition function. This effect is also illustrated in Fig.~\ref{fig2} where we compare Eq.~(\ref{eq35}) at the order $O\left((\kappa_sa)^2\right)$ with the exact MF result of Eq.~(\ref{eq32}). The decrease of the local solvent density with an increase of the bulk dipole concentration at this order results from an intensification of the dielectric screening effect that was shown to be induced by the response of individual dipoles to the surface charge. This reduction of the polarization density results in turn in a decrease of the amplitude of the polarization field at the next order $O\left((\kappa_sa)^4\right)$ in Eq.~(\ref{eq294}). These hierarchical dipolar response relations can be extrapolated to higher orders in solvent concentration by comparing Eq.~(\ref{eq35}) with Eqs.~(\ref{eq294})-(\ref{recP}). It follows that the modification of the single dipole density by collective effects up to the order $O\left((\kappa_sa)^{2n}\right)$ is induced by the response of individual dipoles to their own reaction field at to the order $O\left((\kappa_sa)^{2n-2}\right)$, and this in turn results in a modification of the polarization field at the next order $O\left((\kappa_sa)^{2n+2}\right)$. We note that this picture reminds the ionic correlation effects in inhomogeneous Coulomb liquids characterized by a mutual adjustment of the local ion densities and the electrostatic potential with increasing coupling parameter~\cite{netzcoun,jcp}.

In the opposite regime of concentrated solvents $\kappa_sa\gg1$, the short distance asymptotic limit $z/a\ll1$ of Eq.~(\ref{eq32}) is given by an exponential decay law associated with the characteristic decay length $\kappa_s^{-1}$,
\be\label{eq37}
\delta k_{sp}(z)\simeq\frac{1}{\kappa_s\mu_s}e^{-\kappa_sz}.
\ee
This limiting law, shown in Fig.~\ref{fig2} to perfectly match the prediction of Eq.~(\ref{eq32}) for $\rho_{sb}\geq 4$ M explains the strong reduction of the thickness of the interfacial solvent layer with an increase of the bulk solvent density. The exponential decay indicates that at distances smaller than the
dipole size, the solvent molecules lead exclusively to a \textit{charge screening}, i.e. they screen the induction field $D(z)$ as a concentrated salt solution.
\begin{figure}
\includegraphics[width=1.\linewidth]{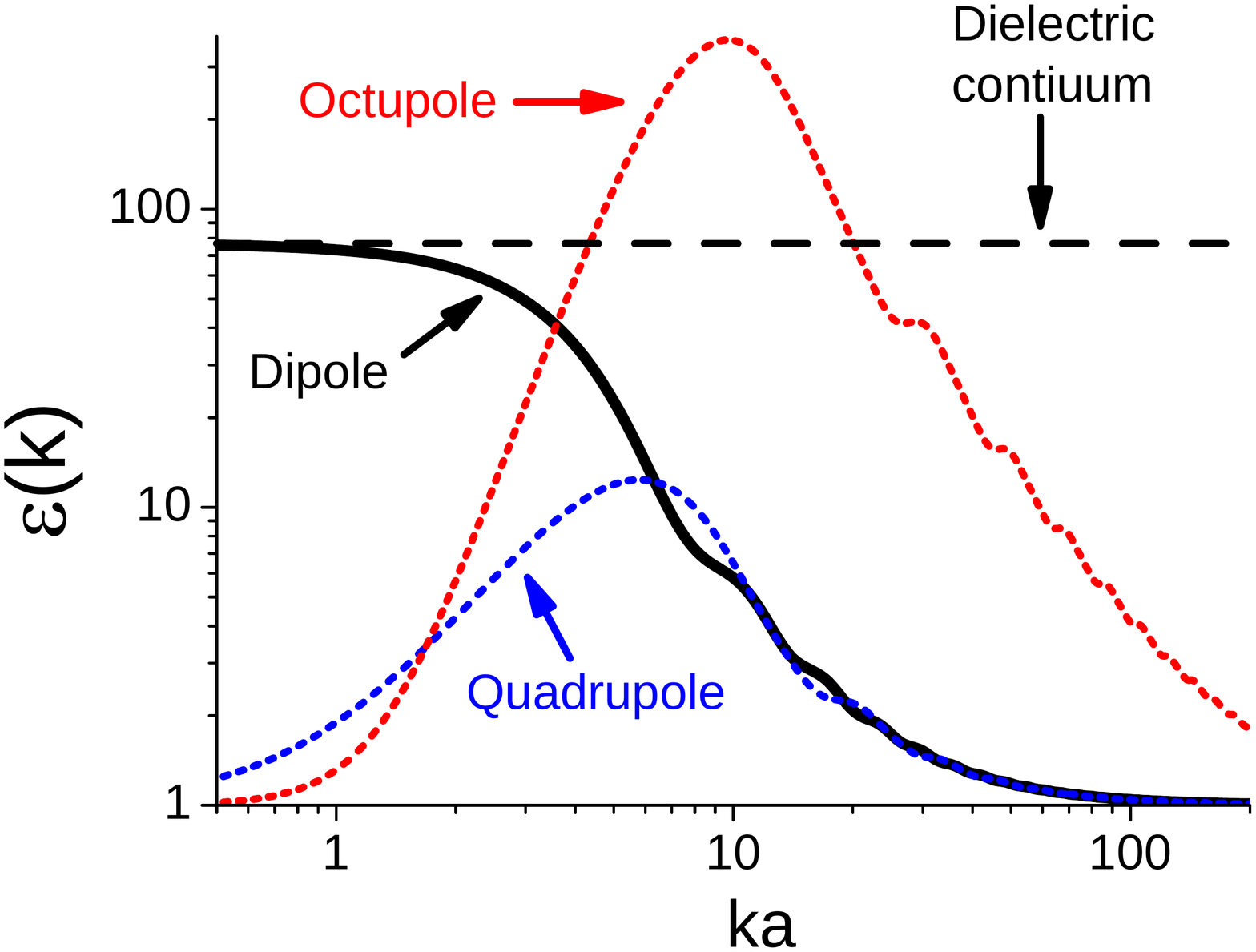}
\caption{(Color online) Fourier transformed dielectric permittivity function Eq.~(\ref{eq28}) for dipoles (solid black curve), quadrupoles (dotted blue curve) and octupoles (dotted red curve) for solvent concentration $\rho_{sb}=55$ M, and molecular size $a=1.0$ {\AA} for dipoles,  $a=0.57$ {\AA} for quadrupoles, and  $a=2.68$ {\AA} for octupoles. The dashed black curve marks the dielectric continuum limit $\te(ka\to0)=\e_w=76.75$ for the dipolar liquid.}
\label{fig3}
\end{figure}

\subsection{Effective dielectric permittivity and ion densities}
\label{efper}

\begin{figure*}
(a)\includegraphics[width=0.45\linewidth]{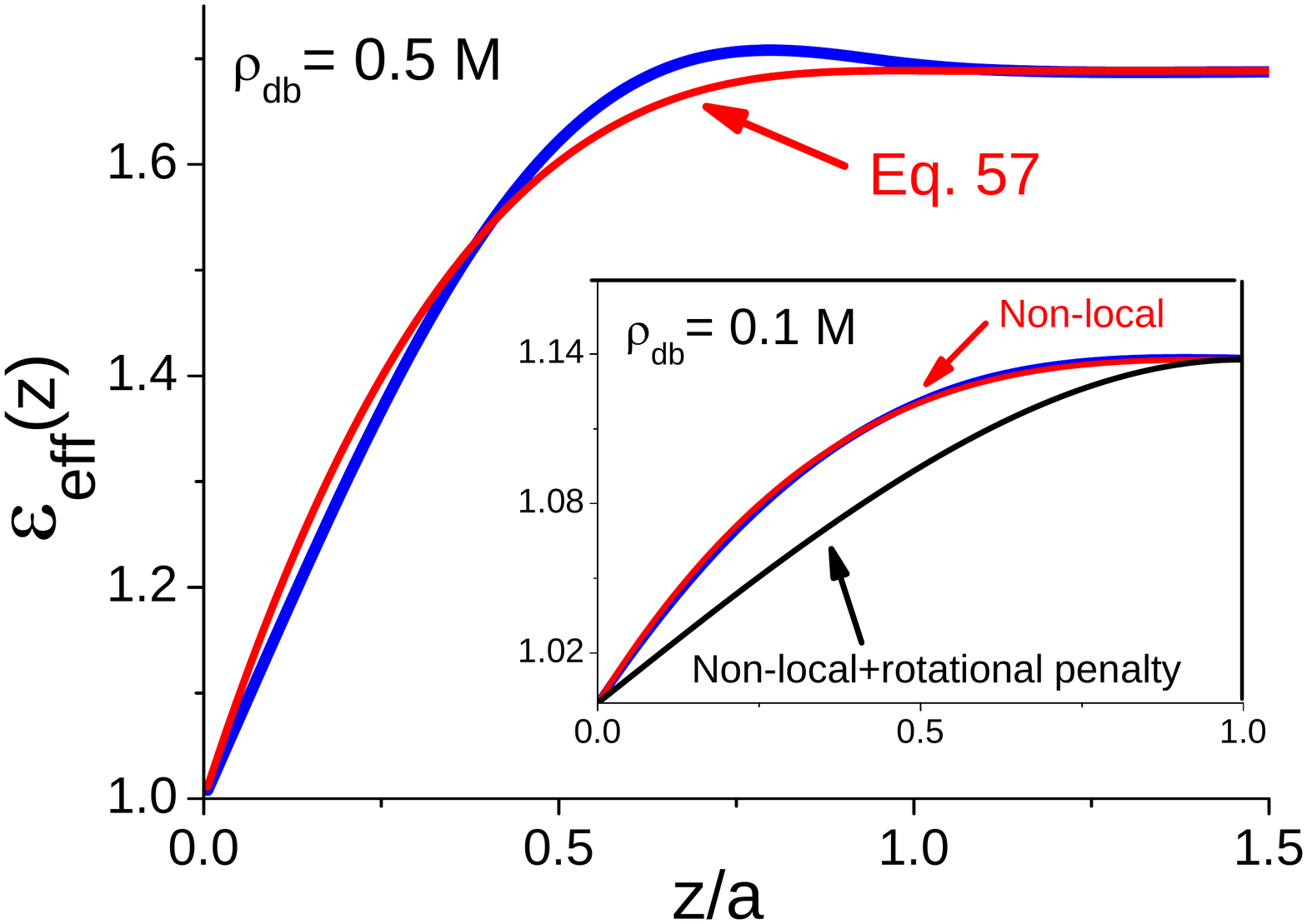}
(b)\includegraphics[width=0.45\linewidth]{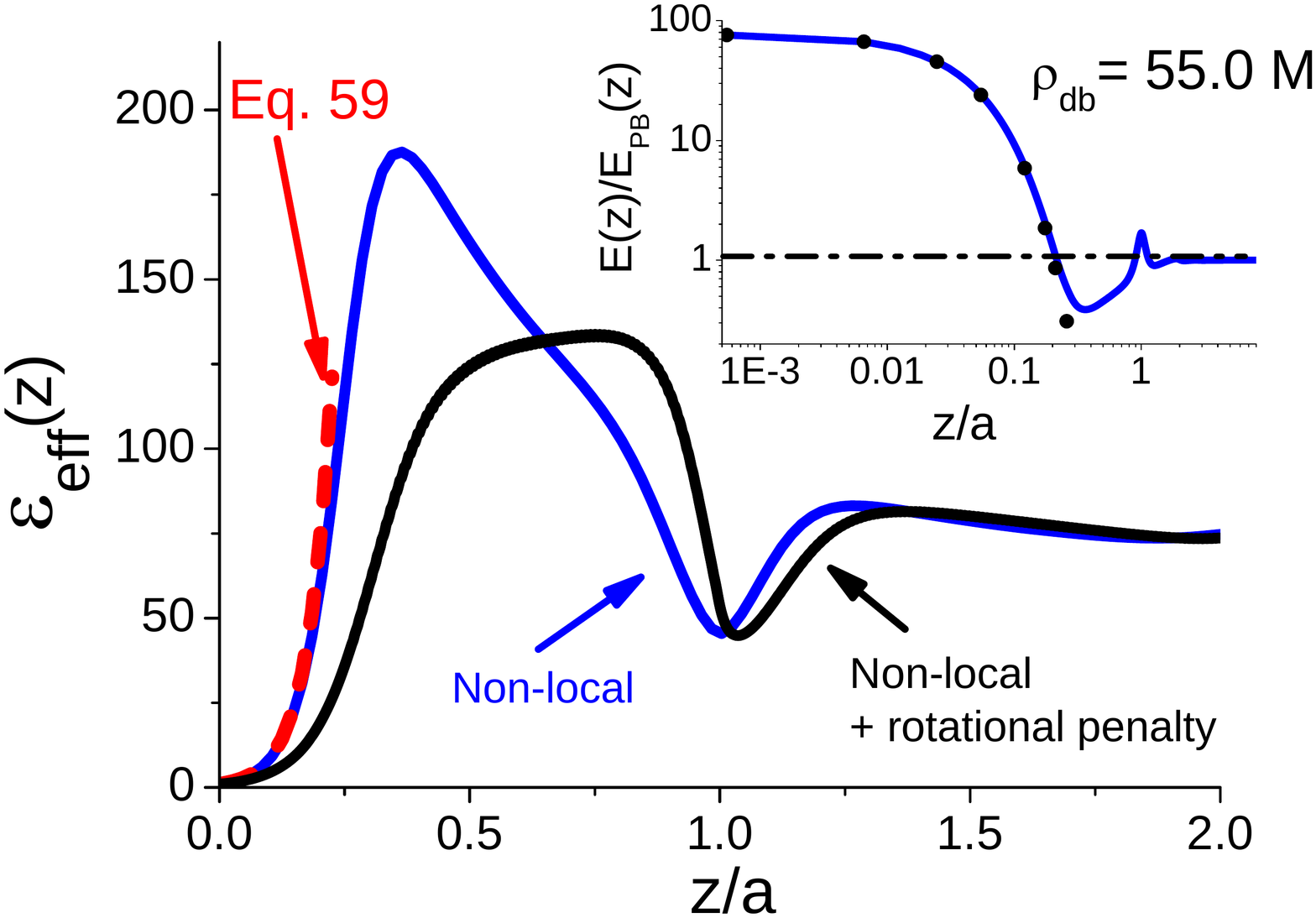}
(c)\includegraphics[width=0.45\linewidth]{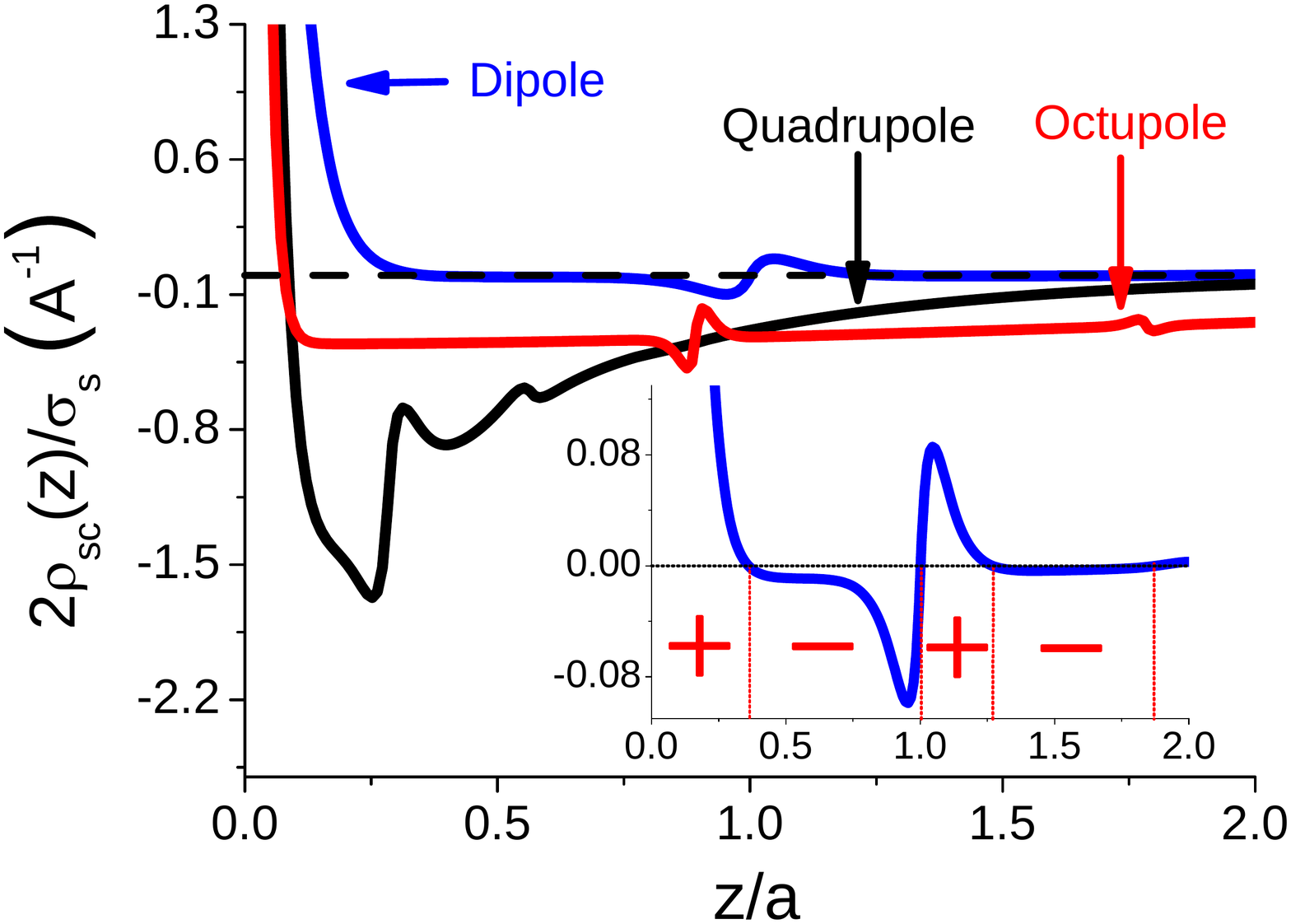}
(d)\includegraphics[width=0.45\linewidth]{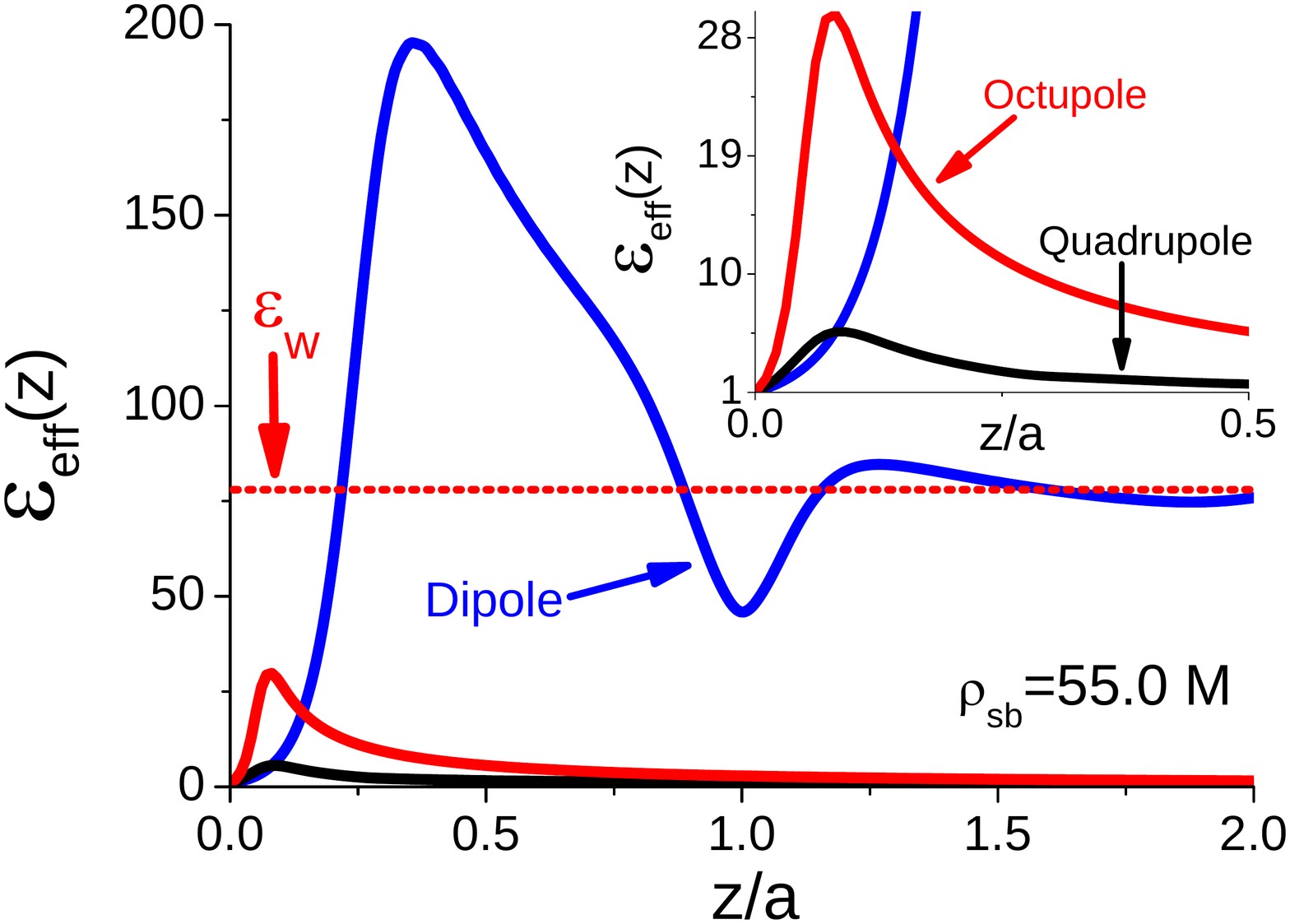}
\caption{(Color online)  Effective dielectric permittivity for the dipolar liquid at the bulk solvent concentrations (a) $\rho_{sb}=0.5$ M (Inset : $\rho_{sb}=0.1$ M), and (b) $\rho_{sb}=55.0$ M (Inset : rescaled electrostatic field from Eq.~(\ref{eq13}) (blue curve) and Eq.~(\ref{eq43}) (black dots)). In the main plots of (a) and (b), solid blue curves are from Eq.~(\ref{eq23}), solid red curves in (a) mark the dilute solvent expression in Eq.~(\ref{eq38}), and the red dashed curve in (b) display the dipolar salt screening regime of Eq.~(\ref{eq42}). The black solid curves in (a) (inset) and (b) account for the rotational penalty for dipoles at a rigid interface. (c) Solvent charge density for the same model parameters as in (b). (d) Comparison of effective permittivity profiles for dipoles (blue curves), quadrupoles (black curves), and octupoles (red curves). In all plots, the solvent concentration is $\rho_{sb}=55.0$ M, the solvent molecular size $a=1.0$ {\AA} for dipoles,  $a=0.57$ {\AA} for quadrupoles, and  $a=2.68$ {\AA} for octupoles.}
\label{fig4}
\end{figure*}

Before considering the variations of the effective dielectric permittivity in real space, it is instructive to understand the two opposite limits of the Fourier transformed permittivity function in Eq.~(\ref{eq28}). As displayed in Fig.~\ref{fig3} for the biological solvent density $\rho_{db}=55$ M, in the IR limit $ka\to0$ corresponding to distances much larger than the molecular size, the function~(\ref{eq28}) tends to the bulk permittivity given by the Debye-Langevin relation, i.e. $\te(k)\to\e_w$. In the opposite ultraviolet (UV) limit $ka\to\infty$ corresponding to the close vicinity of the charge source, the permittivity function tends to the permittivity of the air $\te(k)\to1$. \textcolor{black}{It is interesting to note that the overall shape of this permittivity function resembles the form of the phenomenological Inkson dielectric model~\cite{Ink}.} We will investigate below the corresponding  behavior of the effective permittivity in real space.

The effective dielectric permittivity profile Eq.~(\ref{eq23}) is displayed in Fig.~\ref{fig4} for various solvent concentrations from the dilute to the physiological concentration regime. In the dilute solvent regime, by expanding Eq.~(\ref{eq23}) at the order $O(\left(\kappa_sa\right)^2)$, one obtains for the dielectric permittivity profile the close form expression
\be\label{eq38}
\e_{eff}(z)\simeq1+\frac{(\kappa_sa)^2}{6}\left\{1-\left(1-\frac{z}{a}\right)^3\theta(a-z)\right\}.
\ee
The limiting law~(\ref{eq38}) is reported in the inset of Fig.~\ref{fig4}.a (red curve) for $\rho_{sb}=0.1$ M.  It is seen that the permittivity increases from the dielectric permittivity of the air to the bulk permittivity $\e_w$ in a monotonical way over one molecular size. This results from the interfacial solvent charge formation driven by the single dipole-surface charge attraction in the air medium. Slightly increasing the solvent concentration to $\rho_{sb}=0.5$ M (main plot), the permittivity curve acquires an oscillatory shape around the limiting law~(\ref{eq38}), and exhibits a peak corresponding to a local dielectric increment at $a/2\lesssim z\lesssim a$. This is the density regime where the collective dielectric response mechanism discussed in the previous parts comes into play. In Fig.~\ref{fig4}.b, it is shown that in a polar liquid at the physiological concentration $\rho_{sb}=55$ M, the oscillatory shape of the dielectric permittivity profile becomes more pronounced, with the apparition of alternating dielectric increment and decrement layers characterized by a quasiperiodicity of the order $a$. Furthermore, at the position of the first peak, the local dielectric permittivity exceeds the bulk permittivity almost by a factor 2. By adding into this picture the rigidity of the interface (see Appendix~\ref{rigid} for details), we found that the first dielectric increment peak is significantly decreased, and the permittivity curve is shifted towards larger distances. This results from the reduction of the polarization field induced by the rotational penalty in the region $z<a$.

The oscillations of the background permittivity around $\e_w$ can be shown to result from the formation of successive hydration layers around the charged surface at $z=0$. To this end, we first note that the derivative of the relation~(\ref{eq27}) can be written as
\be\label{eq41}
\frac{d}{dz}\frac{\e_\mathrm{eff}(z)-1}{\e_\mathrm{eff}(z)}=\frac{2\rho_{sc}(z)}{\sigma_s},
\ee
where the solvent charge density is given by Eqs.~(\ref{eq32}) and~(\ref{cp}). According to the relation~(\ref{eq41}), any reversal in the trend of the dielectric permittivity $\e_\mathrm{eff}(z)$ (i.e. any minima or maxima) should originate from an alternation of the sign of the local solvent charge density. This effect is illustrated in the inset of Fig.~\ref{fig4}(c) where we display the renormalized solvent charge density against the separation distance from the surface. In agreement with the number of maxima and minima in the permittivity curve of Fig.~\ref{fig4}(b), there exists four solvation layers of alternating charge over a region of two molecular sizes.

\begin{figure}
\includegraphics[width=1.1\linewidth]{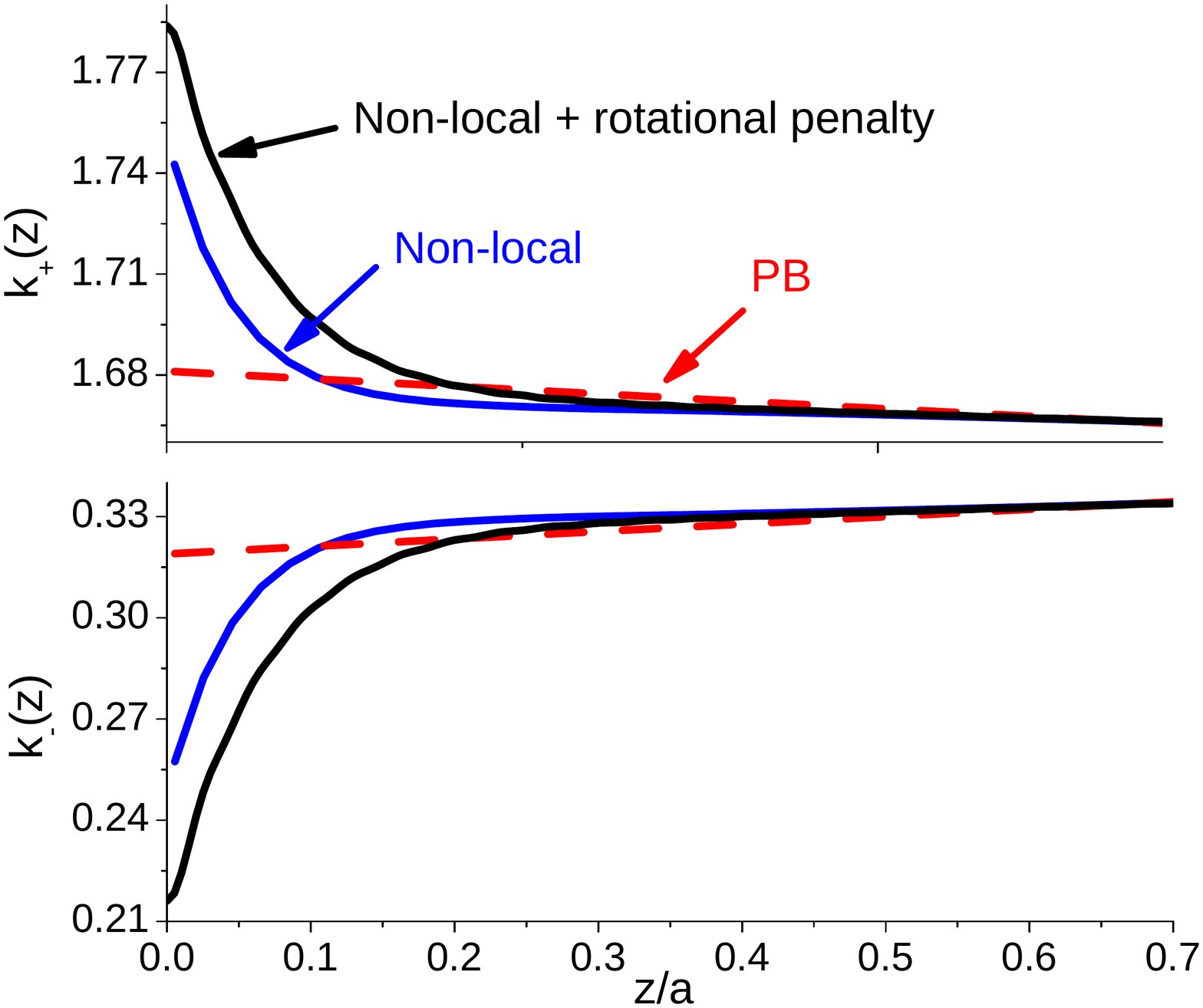}
\caption{(Color online) Counterion (top) and coion (bottom) density profiles of non-polarizable monovalent ions at a penetrable (solid blue curve) and rigid interface (solid black curve). The dashed red curve displays ion densities from the linear PB formalism. The model parameters are $\rho_{ib}=0.01$ M and $\sigma_s=0.05$ $\mbox{e nm}^{-2}$. }
\label{fig5}
\end{figure}

We illustrate in Fig.~\ref{fig5} the ion number densities $k_\pm(z)=1\mp q_i\phi_0(z)$ for monovalent ions $q_i=1$. It is seen that the oscillatory shape of the effective dielectric permittivity in Fig.~\ref{fig4}.b results in weak oscillations of the ion densities around the PB result $k_\pm(z)=1\mp e^{-\kappa_iz}/(q_i\mu_i)$. Furthermore, the surface dielectric deficiency that increases the surface potential amplifies the interfacial counterion attraction and coion repulsion of the PB theory. We also show that accounting for the surface rigidity, the larger dielectric void in the close neighborhood of the interface in Fig.~(\ref{fig4})(b) leads to an amplification of the counterion attraction and coion repulsion. However, the qualitative behavior of ion densities is not modified by the rigidity of the surface.

\textcolor{black}{At this stage, it should be emphasized that despite the simplicity of our linear dipole model and the linear MF approximation, the permittivity curve in Fig.~\ref{fig4}(b) is able to reproduce qualitatively the shape and the periodicity of the transverse dielectric permittivity profiles~\cite{rem} obtained in MD simulations of polar liquids at planar interfaces~\cite{prlnetz,langnetz}. Furthermore, the overall trend of the same dielectric permittivity curve is also in line with the result of AFM experiments for water at charged mica surfaces, where a rise of the local dielectric permittivity  profile from $\e(z)=4$ to $\e_w$ with increasing distance from the charged surface was observed~\cite{expdiel}. This indicates that the consideration of the finite size of solvent molecules in our microscopic polar liquid model is a key improvement over the local theories in order to capture the non-local dielectric response behavior of water in real systems.}

It was shown in the previous part~\ref{sd} that in the concentrated solvent regime $\kappa_sa\gg1$ (i.e. $\rho_{sb}\gg0.1$ M) and at separation distances smaller than the dipole size $z/a\ll1$, the solvent molecules interact with the induction field as a strong salt solution. By computing the short distance limit of Eq.~(\ref{eq23}), one finds that this salt screening translates into an exponentially growing effective dielectric permittivity,
\be\label{eq42}
\e_{eff}(z)\simeq e^{\kappa_sz}.
\ee
This limiting law is displayed in Fig.~\ref{fig4}(b) by the dashed red curve. Substituting now Eq.~(\ref{eq42}) into the relation~(\ref{eq22}), one obtains for the electrostatic field
\be\label{eq43}
E(z)=\frac{\ew}{q_i\mu_i}e^{-\kappa_sz}.
\ee
We illustrate in the inset of Figs.~\ref{fig4}.c the electrostatic field Eq.~(\ref{eq13}) (blue curve) and its asymptotic limit Eq.~(\ref{eq43}) (black dotes) rescaled by the field of the PB formulation $E_\mathrm{PB}(z)=e^{-\kappa_iz}/(q_i\mu_i)$. First of all, one notices in this figure and Eq.~(\ref{eq43}) that the PB formalism that cannot account for the dielectric screening deficiency on the surface underestimates the surface field by a factor $\e_w$. Then, we see that the fast drop of the electrostatic field to the order of magnitude of $E_\mathrm{PB}(z)$ is solely driven by the dipolar charge screening. This means that at physiological solvent concentrations, the interfacial decay of the surface field is induced by the dipolar salt screening, rather than the dielectric screening resulting from the preferential orientation of dipoles. However, we note that our MF level of approximation neglects image-dipole interactions, which are expected to weaken the dipolar salt screening effect.

\subsection{Multipolar contributions to non-local dielectric response}
\label{multi}

We investigate in this part the contribution of the multipolar moments of water to the dielectric permittivity of the liquid. To this end, we incorporated the relative multipole/dipole moments of the TIP4P/2005 water site model into the present theory by computing the ratios between the dipolar moment $\mu_0=Qa$, and the quadrupolar and octupolar moments $\Theta_0=Qa^2/4$ and $\Omega_0=-Qa^3/27$ of the solvent molecules depicted in Fig.~\ref{fig6}(a)-(c). By setting these ratios to the rescaled multipole moments $\Theta_0/\mu_0=0.08$ {\AA} and $\Omega_0/\mu_0=-0.71$ $\mathrm{{\AA}}^2$ given in Ref.~\cite{niu} for the same linear multipole geometries as in our model, we obtained the corresponding molecular sizes.  For multipolar solvent molecules, the permittivity function~(\ref{eq28}) should be evaluated with the multipolar charge structure factors and molecular sizes given by $F(\tk)=3-8\sin(\tk/2)/\tk+\sin(\tk)/\tk$ and $a=0.57$ {\AA} for quadrupolar molecules, and $F(\tk)=10-45\sin(\tk/3)/\tk+9\sin(2\tk/3)/\tk-\sin(\tk)/\tk$ and $a=2.68$ {\AA} for octupoles, where we introduced the adimensional wave vector $\tk=ka$.

The shape of the Fourier transformed dielectric permittivity profiles are displayed in Fig.~\ref{fig3} for dipoles, quadrupoles, and octupoles. It is seen that the behavior of $\te(k)$ is qualitatively similar for quadrupoles and octupoles. Namely, the permittivities tend to the air permittivity for large wave vectors as in the dipolar case, which is associated with the dielectric void in the neighborhood of the charged surface. However, unlike the dipolar permittivity curve, both functions exhibit a peak corresponding to a region of maximum dielectric screening in real space, and converge again towards the air permittivity for $ka\to0$. The latter aspect stems clearly from the zero dipolar moment of quadrupolar and octupolar molecules. As it will be shown next, this unables them to polarize the medium at large distances from the charge sources.

The effective dielectric permittivity profiles associated with quadrupolar and octupolar molecules are displayed in main plot and the inset of Fig.~\ref{fig4}(d). In agreement with the wave vector dependence of the Fourier transformed permittivity functions in Fig.~\ref{fig3}, the dielectric permittivities tend to the air permittivity at the charged surface and in the bulk limit, with a maximum dielectric screening peak in between. This aspect can be explained in an intuitive way in terms of the solvent charge density that we display in Fig.~\ref{fig4}(c). In this plot, one first notices the strong oscillatory behavior of solvent charge densities, characterized by sharp kinks located at the separation distances between the elementary charges on the solvent molecules. Then, for the multipolar liquids, one notices the large amplitude of the second negative solvation shell following the first positive one at the interface. Hence, unlike the dipolar solvent molecules that result in a net positive accumulated charge in the interfacial area, the first positive charge layer of the quadrupolar and octupolar liquids are almost exactly canceled by the next negative solvation layer. According to Eq.~(\ref{eq27}), this results in a vanishing multipolar contribution to the bulk effective permittivity of the medium. Finally, we show in the main plot of Fig.~\ref{fig4}(d) that even in the interfacial region, the background dielectric permittivity induced by dipoles largely dominate the multipolar one. This observation is in line with recent MD simulations where the multipolar moments of water molecules were shown to weakly affect the transverse permittivity of the polar liquid~\cite{langnetz}.

\subsection{Ionic polarizability}
\label{pol}

This part is devoted to the effect of ionic polarizability on ionic partitions and the dielectric propreties of a dipolar liquid. The charge composition of polarizable ions of two species  with an equal bulk concentration $\rho_{bi}$  and electronic cloud radius $b_p$ is illustrated in Fig.~\ref{fig6}(d). The  elementary charges on the molecules have valency $e_{\pm}=\mp1$ and $c_{\pm}=\pm 2$, with the subscripts $\pm$ denoting the overall positive and negative molecules in Fig.~\ref{fig6}(d).

For this ionic charge geometry, the susceptibility function introduced in Eq.~(\ref{eq16}) takes the form
\bea\label{eq54}
\chi(z)&=&\frac{p_0^2\rho_{sb}}{2a}\left(1-\frac{|z|}{a}\right)^2\theta\left(a-|z|\right)\\
&&+8\rho_{ib}b_p\left\{\frac{1}{\sqrt\pi}\exp\left(-\frac{|z|^2}{4b_p^2}\right)-\frac{|z|}{2b_{p}}\mathrm{Erfc}\left(\frac{|z|}{2b_{p}}\right)\right\}\nonumber.
\eea
We note that the function in the bracket on the rhs of Eq.~(\ref{eq54}) behaves as $\sim \tilde{z}^{-2}e^{-\tilde{z}^2/4}$ for $\tilde{z}=z/b_p\gg1$. This indicates that the induced ionic polarizability extends the range of the liquid polarizability beyond the solvent molecular size $a$, with a fast decay that obeys a gaussian law characterized by the decay length $b_p$. Furthermore, for the same charge geometry, the Fourier transformed dielectric permittivity in Eq.~(\ref{eq11}) takes the form
\bea\label{eq55}
\te(k)&=&1+\frac{\kappa_d^2}{k^2}\left[1-\frac{\sin(ka)}{ka}\right]+\frac{32\pi\ell_B\rho_{ib}}{k^2}\left(1-e^{-b_p^2k^2}\right).\nonumber\\
\eea
The bulk dielectric permittivity that follows from the IR limit of Eq.~(\ref{eq55}) is given by $\e_w=1+4\pi\ell_Bp_0^2\rho_{sb}/3+32\pi\ell_Bb_p^2\rho_{ib}$. Identifying the ionic polarizability as $\alpha=4b_p^2$, one finds that the IR limit of Eq.~(\ref{eq55}) yields the correction from the induced polarizability to the bulk permittivity derived in Ref.~\cite{netzvdw}.

\begin{figure*}
\includegraphics[width=0.45\linewidth]{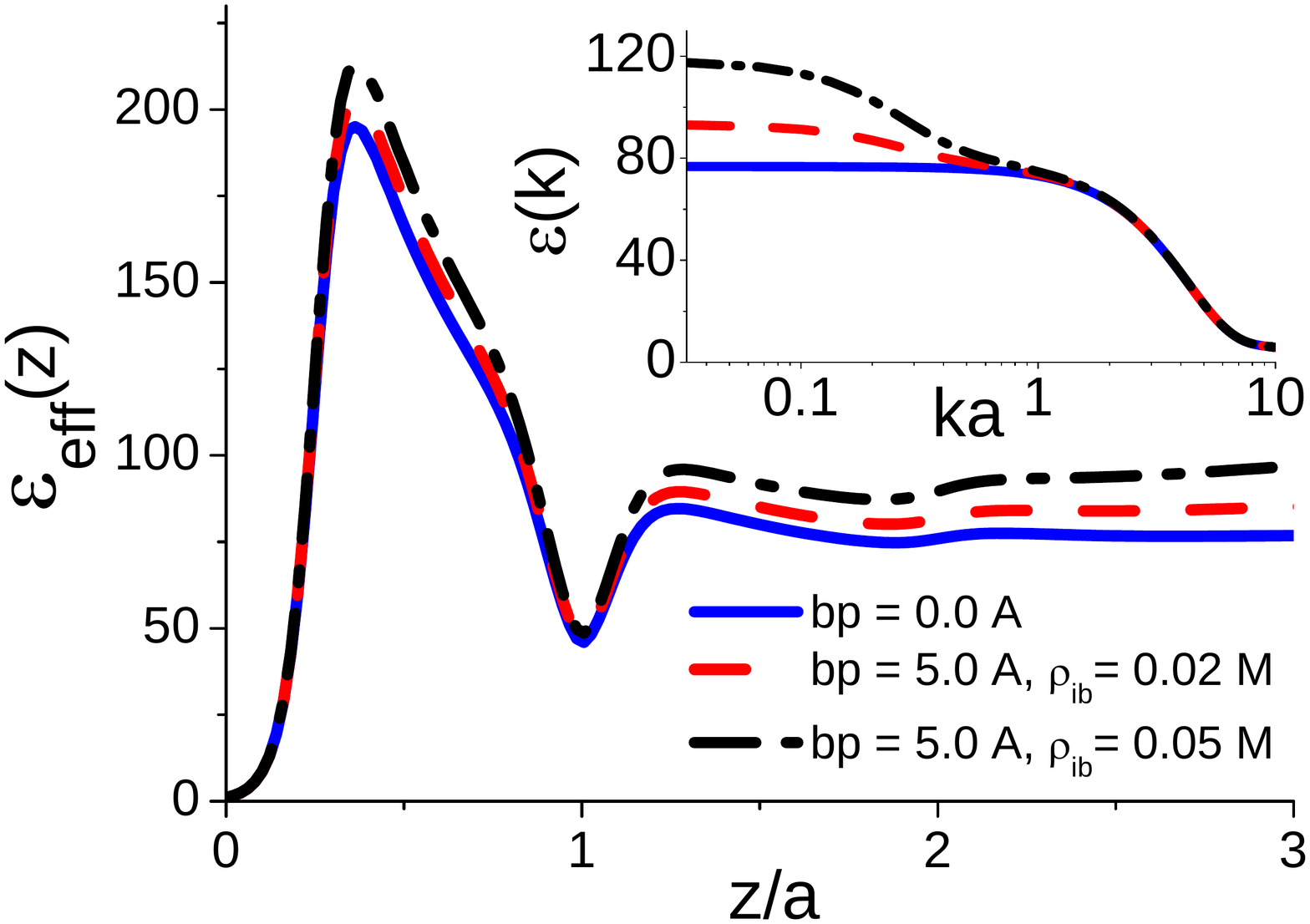}
\includegraphics[width=0.5\linewidth]{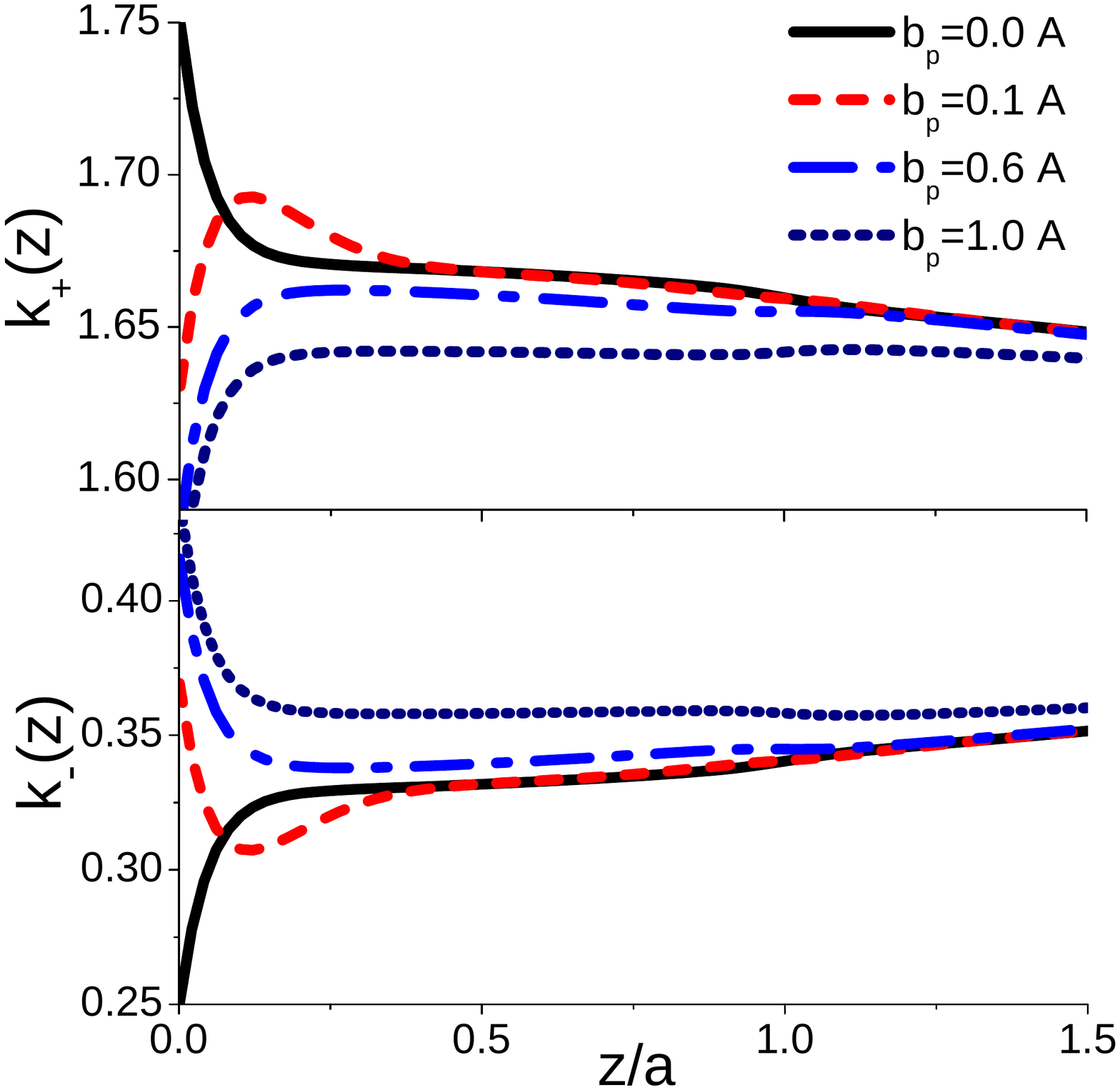}
\caption{(Color online) (a) Effect of induced ion polarizability on the dielectric permittivity in real space (main plot) and Fourier space (inset). (b) Counterion (top) and coion (bottom) densities of polarizable ions for the model parameters $\rho_{ib}=0.01$ M, and $\sigma_s=0.05$ $\mbox{e nm}^{-2}$.}
\label{fig9}
\end{figure*}

The behavior of the function~(\ref{eq55}) is illustrated in the inset of Fig.~\ref{fig9}(a) at the solvent concentration $\rho_{sb}=55.0$ M, and two values of the bulk ion concentration. In order to clearly illustrate the contribution from the induced polarizability, a considerably large value $b_p=5$ {\AA} was chosen. It is seen that the ionic polarizability affects the permittivity function mainly at small wavelengths. More precisely, for finite polarizability with $b_p>a$, the permittivity function exhibits a bimodal decay at the wavelengths corresponding to the average fluctuations of the electronic cloud radius $k_1\sim b_p^{-1}$ and the solvent size $k_2\sim a^{-1}$. The corresponding behavior of the effective permittivity in real space is also shown in the main plot of Fig.~\ref{fig9}(a). In agreement with the trend of the function $\te(k)$ in Fourier space, the local dielectric permittivity increases with the induced polarizability,

In the weak surface charge regime and for the charge composition depicted in Fig.~\ref{fig6}(d), the number density partition function of polarizable ions follows from Eq.~(\ref{ipar}) as
\be\label{eq56}
\delta k_{\pm p}(z)=4\ell_w\sigma_s\int_0^\infty\frac{\mathrm{d}k\cos(kz)}{\kappa_i^2+k^2\te(k)/\ew}\left(1\pm 2 e^{-b_p^2k^2}\right).
\ee
This density profile is displayed in Fig.~\ref{fig9}(b) for different values of the induced polarizability $b_p$. One sees that in the presence of a finite polarizability, the trend of the densities in the interfacial area are completely reversed. Namely, for weak polarizabilities $b_p<a$, the coion density reaches a minimum, and starts to increase towards the surface with decreasing separation distance from the interface, while counterion density exhibits a concentration peak at a characteristic distance, and decreases towards the interface. For ionic polarizabilities close to the solvent molecular size $b_p\simeq a$, the interfacial reduction of the counterion attraction and coion depletion becomes monotonous.

The surface propensity of coions and exclusion of counterions with finite polarizability in Fig.~\ref{fig9}(b) clearly results from their discrete charge structure. Indeed, the ionic polarizability favors the interaction of the negative (positive) charge $e_i$ on the counterion (coion) with the surface charge. This result in a reversal of the ion partition trends at the interface. Hence, unlike the effective permittivity of the liquid in Fig.~\ref{fig9}(a), the ion densities are substantially affected by the induced polarizability. This result disagrees with the conclusion of Ref.~\cite{frydel} where the consideration of the ionic polarizability in the point dipole limit was shown to weakly affect the ion densities in the weak electrostatic coupling regime. This shows that our proper treatment of ionic polarizability by explicitly considering the extended charge structure of ions is crucial.

\section{Summary and Conclusions}

In conclusion, we have presented a microscopic theory of non-local electrostatic interactions in polar liquids. \textcolor{black}{It was shown that unlike previous approaches treating the solvent molecules as point dipoles~\cite{dunyuk,orland1,orland2,epl}, our formulation accounting for the finite size of solvent molecules can qualitatively capture the non-local dielectric response of polar liquids at charged interfaces.}

In the first part of the article, we derived the field theory of the polar liquid composed of linear multipoles of finite size, and containing polarizable ions  modeled as Drude oscillators. From the saddle point solution of the partition function, we obtained  a non-local Poisson-Boltzmann (NLPB) equation. In the rest of the article, we investigated the non-local electrostatic interactions embodied in this equation within \textcolor{black}{the linear dielectric response regime for polar liquids and ions in contact with a weakly charged planar interface.}

In the second part of the article, we introduced a mapping from the microscopic model to the macroscopic formulation of non-local electrostatics. A key result of this part is the expression~(\ref{eq27}) for the background dielectric permittivity of the medium in terms of the accumulated polarization charge between the charged interface and the liquid. In agreement with MD simulations~\cite{Hans,prlnetz,langnetz} and AFM experiments of water at charged surfaces~\cite{expdiel}, this relation predicts an interfacial layer associated with a reduced dielectric permittivity, and resulting from the reduction of the polarization field towards the interface.

Then, in the third part, we thoroughly analyzed the dipolar correlations in the solvent model. We found that the non-local dielectric response of the liquid to charge sources is driven by a cooperative mechanism resulting from the response of the solvent molecules to their own polarization field. We also showed that our model can qualitatively reproduce the shape and the periodicity of the transverse dielectric permittivity profiles obtained in MD simulations of water at charged interfaces~\cite{prlnetz,langnetz}. The fluctuations of the dielectric permittivity around the bulk one was shown to result from the formation of successive hydration layers of alternating net charge in the interfacial region. At the next step, we evaluated the contribution of the multipolar moments of water to the dielectric permittivity of the medium, and found that the multipolar contributions are largely dominated by the dipolar one. This observation is in line with the evaluation of multipolar contributions to the transverse dielectric permittivity in MD simulations~\cite{prlnetz,langnetz}.

Finally, we investigated the effect of the induced ion polarizability on the ionic partitions at a charged interface. It was shown that in the presence of an arbitrary surface charge, the polarizability completely reverses the interfacial ion density predictions of the PB approach, resulting in a surface propensity of coions and depletion of counterions. This result disagrees with previous works based on the point dipole approximation, where a perturbative correction from the induced polarizability to ion densities was observed~\cite{frydel}. This indicates that the consideration of the extended charge structure of polarizable molecules is crucial.

Being a first microscopic theory of non-local electrostatic interactions, the NLPB approach possess limitations. First of all, the present theory neglects excluded volume effects associated with solvent molecules and ions. This complication could be incorporated into the theory by imposing a steric Fermi distribution to particles~\cite{bor} or modeling the hard-core repulsions between them with a Yukawa potential as in Refs.~\cite{duny,jstat,jcp1}. Furthermore, we focused in the present work exclusively on \textcolor{black}{the linear dielectric response regime of liquids in contact with a weakly charged single interface}. We wish to consider the non-linear effects embodied in Eq.~(\ref{eq1}) for polar liquids in confined geometries in an upcoming article. Then, our discussion on the non-local dielectric response of the liquid was based on the MF formulation. However, it should be noted that through the field theoretic formulation of the multipolar liquid model in Eq.~(\ref{HamFunc}), the present work sets non-local electrostatics on a solid theoretical framework. Accompanied with MC simulations of the solvent model Eq.~(\ref{HamFunc}) at interfaces, this consistent framework will allow to consider in the future non-local electrostatic correlations in inhomogeneous polar liquids in a systematic way. Finally, the linear multipole model can be easily generalized to different water site models used in MD simulations. These extensions will allow direct comparisons of the theory with MD simulation results of biological and interfacial systems with explicit water.
\\
\acknowledgements  This work has been in part supported by The Academy of Finland through its Centres of Excellence Program (project no. 251748) and NanoFluid grants.
\smallskip
\appendix
\section{Low density expansion}
\label{dil}

This appendix is devoted to the evaluation of single dipole densities in contact with a plane located at $z=0$ and carrying a surface charge $\sigma_s$. In the dilute liquid regime, the expansion of the adimensional grand potential $\Omega_G=-\ln Z_G$  in powers of the particle fugacities yields the grand potential in the form
\be~\label{sin}
\Omega_{1p}=\Omega_0+\lan H-H_0\ran_0,
\ee
where the statistical average is taken with the gaussian Hamiltonian
\be\label{eq01}
H_0[\phi]=\int\mathrm{d}\br\left[\frac{\left[\nabla\phi(\br)\right]^2}{8\pi\ell_B(\br)}-i\sigma(\br)\phi(\br)\right],
\ee
Furthermore, the gaussian part of the grand potential in Eq.~(\ref{sin}) is given by $\Omega_0=-\ln Z_0$, and the reference partition function reads
\bea\label{eq02}
Z_0&=&\int \mathcal{D}\phi\;e^{-H_0[\phi]}\\
&=&\sqrt{\mathrm{det}(v_c)}\exp\left[\int\frac{\mathrm{d}\br\mathrm{d}\br'}{2}\sigma(\br)v_c(\br-\br')\sigma(\br')\right].\nonumber
\eea
Evaluating the field theoretic averages in Eq.~(\ref{sin}), one obtains the grand potential in the form
\bea\label{eq03}
\Omega_{1p}&=&\Omega_0-\sum_i\lambda_i\int\mathrm{d}\br e^{-W_i(\br)-\psi_i(\br)}\\
&&-\Lambda_se^{Q^2\frac{\ell_B}{a}}\int\frac{\mathrm{d}\br\mathrm{d}\bom}{4\pi}e^{-W_{s1}(\br)-W_{s2}(\br,\br+\ba)-\psi_s(\br,\ba)},\nonumber
\eea
where we introduced respectively the ionic and dipolar potential of mean forces (PMFs)
\bea\label{eq04}
\psi_i(\br)&=&\int\mathrm{d}\br'\mathrm{d}\br''\sigma(\br')v_c(\br'-\br'')q_i\delta(\br''-\br)\\
\label{eq05}
\psi_s(\br,\ba)&=&\int\mathrm{d}\br'\mathrm{d}\br''\sigma(\br')v_c(\br'-\br'')Q\left[\delta(\br''-\br)\right.\\
&&\hspace{4.2cm}\left.-\delta(\br''-\br-\ba)\right].\nonumber
\eea
We recognize in Eq.~(\ref{eq05}) the coupling potential in the air medium between a single dipole and the fixed surface charge. Evaluating the integrals in Eq.~(\ref{eq05}) with the Coulomb potential
\be\label{eq07}
v_c(\br-\br')=4\pi\ell_B\int\frac{\mathrm{d}^3\bk}{(2\pi)^3}\frac{e^{i\bk\cdot(\br-\br')}}{k^2},
\ee
one obtains the dipolar PMF in the form
\be\label{eq33}
\psi_s(z,a_z)=\frac{|z|}{\mu_s}-\frac{|z+a_z|}{\mu_s}.
\ee
We notice that Eq.~(\ref{eq33}) is simply the 1D interaction potential of a finite size dipole in the air medium  with a constant electric field $E_z=1/(Q\mu_s)=2\pi\ell_B\sigma_s$.

We now note that the dipole density is defined as
\bea
\label{eq051}
\rho_s(\br)&=&\frac{\delta\Omega_{1p}}{\delta W_{s1}(\br)}\nonumber\\
&=&\rho_{sb}\int\frac{\mathrm{d}\bom}{4\pi}e^{-W_{s1}(\br)-W_{s2}(\br,\br+\ba)-\psi_s(\br,\ba)},
\eea
where we accounted for the relation between the fugacity and the bulk dipole concentration $\rho_{sb}=\Lambda_se^{Q^2\frac{\ell_B}{a}}$. Setting the dipolar wall potential to zero and expanding Eq.~(\ref{eq051}) at the linear order in $\psi_s$, one gets the dipolar partition function as
\be\label{321}
k_{sp}(z)=1-\int_{-a}^{a}\frac{d a_z}{2a}\psi_s(z,a_z),
\ee
Evaluating the integral in Eq.~(\ref{321}) with the PMF~(\ref{eq33}), one finally obtains the excess dipolar partition function in the form
\be\label{eq34}
\delta k_{sp}(z)=\frac{a}{2\mu_s}\left(1-\frac{z}{a}\right)^2\theta(a-z).
\ee

\section{Evaluation of the electrostatic potential at rigid interfaces}
\label{rigid}

We present in this appendix the evaluation of the electrostatic potential for the polar liquid symmetrically partitioned around a charged rigid interface. The symmetric ion distribution around the interface leads to the vanishing ionic wall potential $W_i(z)=0$, whereas the surface rigidity that restricts the dipolar rotations in the region $|z|<a$ can be taken into account by introducing the dipolar potential $W_s(\br,\bom)=W_d(a,a+a_z)$, with $W_d(z,z+a_z)=0$ if $z(z+a_z)>0$, and $W_d(z,z+a_z))=\infty$ for $z(z+a_z)<0$. In the weak potential approximation, the MF equation~(\ref{eq1}) accounting for this dipolar wall potential reads
\be\label{eq44}
\Delta\phi(z)-\e_w\kappa_i^2\phi(z)+4\pi\ell_B\left[\sigma_s(z)+\rho_{sb} k_{sc}(z)\right]=0,
\ee
with the solvent charge density
\bea\label{eq442}
k_{sc}(z)&=&2Q^2\theta(z)\int_{-\mathrm{min}(a,z)}^{a}\frac{\mathrm{d}a_z}{2a}\left[\phi(z+a_z)-\phi(z)\right]\nonumber\\
&&+2Q^2\theta(-z)\int_{-a}^{\mathrm{min}(a,|z|)}\frac{\mathrm{d}a_z}{2a}\left[\phi(z+a_z)-\phi(z)\right].\nonumber\\
\eea

Because the integral boundaries in Eq.~(\ref{eq442}) depend on the distance from the interface, we cannot solve Eq.~(\ref{eq44}) in Fourier space. We will thus solve this equation by using a perturbative inversion method. To this aim, we reexpress Eq.~(\ref{eq44}) in the form
\bea\label{eq45}
&&\Delta\phi(z)-\e_w\kappa_i^2\phi(z)+\kappa_s^2\int_{-a}^{a}\frac{\mathrm{d}a_z}{2a}\left[\phi(z+a_z)-\phi(z)\right]\nonumber\\
&&=-4\pi\ell_B\left[\sigma_s(z)-\lambda_d\rho_{sb}\delta k_{sc}(z)\right],
\eea
where the excess dipolar charge density reads
\bea\label{eq46}
\delta k_{sc}(z)&=&2Q^2\theta(z)\theta(a-z)\int_{-a}^{-z}\frac{\mathrm{d}a_z}{2a}\left[\phi(z+a_z)-\phi(z)\right]\nonumber\\
&&2Q^2\theta(-z)\theta(a-|z|)\int_{|z|}^{a}\frac{\mathrm{d}a_z}{2a}\left[\phi(z+a_z)-\phi(z)\right].\nonumber\\
\eea
We note that in Eq.~(\ref{eq45}), we introduced the expansion parameter $\lambda_d$ in order to keep track of the perturbative order. With the use of the electrostatic kernel Eq.~(\ref{eq5}) that reads for the dipolar charge distribution
\bea\label{eq462}
G^{-1}(z,z')&=&\frac{-\partial_z^2+\e_w\kappa_i^2}{4\pi\ell_B}\delta(z-z')\\
&&-Q^2\rho_{sb}\int_{-a}^{a}\frac{\mathrm{d}a_z}{2a}\left\{\delta(z'-z-a_z)\right.\nonumber\\
&&\hspace{2cm}\left.+\delta(z'-z+a_z)-2\delta(z'-z)\right\},\nonumber
\eea
one can invert the relation~(\ref{eq45}) and express the potential in the form
\be\label{eq47}
\phi(z)=\phi_0(z)+\delta\phi(z),
\ee
where $\phi_0(z)$ corresponds to the electrostatic potential of Eq.~(\ref{eq12}) for a permeable surface, and the excess potential accounting for the corrections from the rotational penalty reads
\be\label{eq48}
\delta \phi(z)=-\lambda_d\frac{\kappa_s^2}{4\pi\ell_B}\int_{-\infty}^{\infty}\mathrm{d}z'G_0(z,z')\delta k_{sc}(z').
\ee
Furthermore, by expanding the excess potential in powers of the parameter $\lambda_d$,
\be\label{eq49}
\delta\phi(z)=\sum_{n\geq1}\lambda_d^n\phi_n(z),
\ee
substituting this expansion into Eq.~(\ref{eq48}) with Eqs.~(\ref{eq46})-(\ref{eq47}), and identifying the equal powers of $\lambda_d$, one gets the following recurrence relation between the components of the excess potential,
\bea\label{eq50}
\phi_n(z)&=&-\left(\kappa_sa\right)^2\int_0^{a}\frac{\mathrm{d}z'}{a}T(z,z')\\
&&\hspace{1.4cm}\times\int_{z'}^{a}\frac{d a_z}{2a}\left[\phi_{n-1}(z'-a_z)-\phi_{n-1}(z')\right],\nonumber
\eea
where we introduced the function
\be\label{eq51}
T(z,z')=\frac{2}{\pi}\int_0^\infty\mathrm{d}q\frac{\cos(qz/a)\cos(qz'/a)}{\left(\kappa_ia\right)^2\e_w+q^2\te(q)},
\ee
with the rescaled wave vector $q=k/a$. We note that deriving the expression~(\ref{eq50}), we made use of the reflection symmetry of the potential $\phi(-z)=\phi(z)$.

The steric corrections to the electrostatic potential $\phi_n(z)$ associated with the interface rigidity is evaluated numerically from the recurrence relation in Eq.~(\ref{eq50}) for increasing $n$ until numerical convergence is achieved. The dielectric permittivity profile in Fig.~\ref{fig4}(b) (solid black curve) was obtained by injecting the converged result into Eq.~(\ref{eq22}).


\begin{thebibliography}{99}
\bibitem {rem} In MD simulations of polar liquids at planar geometries~\cite{Hans,prlnetz}, the dielectric response of the solvent exhibits an effective permittivity function composed of a longitudinal and a transverse component. This feature stems from the spherical symmetry breaking induced by correlation effects, such as interfacial dipolar alignment driven by image-dipole interactions. Our preliminary results beyond the MF level analysis of the model Eq.~(\ref{HamFunc}) show that the latter indeed embodies this dielectric anisotropy effect. However, this point is beyond the scope of the present article.
\end{thebibliography}
\end{document}